\documentclass[9pt,twocolumn,twoside]{fgm_gsajnl}
\usepackage{chngcntr}
\usepackage{lineno}
\usepackage{wasysym}
\articletype{inv} 

\usepackage{nameref}
\usepackage[user,titleref]{zref}
\renewcommand{\eqref}[1]{Equation~\ref{#1}}

\newcommand{\NumMax}{\mathcal{N}}
\newcommand{\Tr}{\sum_{\tau}}
\newcommand{\Avr}[1]{\left \langle #1 \right \rangle}	
\newcommand{\Measure}[1]{\mathcal{D} #1}

\newcommand{\JointProb}{\mathcal{P}(R_1, R_2, R)}
\newcommand{\ProbBeneficial}{P_b}
\newcommand{\ProbReciprocal}{P_\mathrm{RSE}}
\newcommand{\ProbSimple}{P_{\mathrm{SSE}}}
\newcommand{\ProbMagnitude}{P_\mathrm{ME}}

\newcommand{\ProbReciprocalBeneficial}{P^{b}_{\mathrm{RSE}}}
\newcommand{\ProbSimpleBeneficial}{P^{b}_{\mathrm{SSE}}}

\title{Genotypic complexity of Fisher's geometric model}

\author[$\ast$]{Sungmin Hwang}
\author[$\dagger$,1]{Su-Chan Park}
\author[$\ast$]{Joachim Krug}

\affil[$\ast$]{Institut  f\"ur Theoretische Physik, Universit\"at zu K\"oln, 50937 K\"oln, Germany}
\affil[$\dagger$]{Department of Physics, The Catholic University of Korea, Bucheon 14662, Republic of Korea}


\runningtitle{Fisher's Geometric Model} 

\correspondingauthor{S. Hwang, S.-C. Park, and J. Krug}

\begin{abstract}
Fisher's geometric model was originally introduced to argue that complex adaptations must occur in small steps because of pleiotropic constraints. When supplemented with the assumption of additivity of mutational effects on phenotypic traits, it provides a simple mechanism for the emergence of genotypic epistasis from the nonlinear mapping of phenotypes to fitness. Of particular interest is the occurrence of reciprocal sign epistasis, which is a necessary condition for multipeaked genotypic fitness landscapes. Here we compute the probability that a pair of randomly chosen mutations interacts sign epistatically, which is found to decrease with increasing phenotypic dimension $n$, and varies nonmonotonically with the distance from the phenotypic optimum. We then derive expressions for the mean number of fitness maxima in genotypic landscapes comprised of all combinations of $L$ random mutations. This number increases exponentially with $L$, and the corresponding growth rate is used as a measure of the complexity of the landscape. The dependence of the complexity on the model parameters is found to be surprisingly rich, and three distinct phases characterized by different landscape structures are identified. Our analysis shows that the phenotypic dimension, which is often referred to as phenotypic complexity, does not generally correlate with the complexity of fitness landscapes and that even organisms with a single phenotypic trait can have complex landscapes. Our results further inform the interpretation of experiments where the parameters of Fisher's model have been inferred from data, and help to elucidate which features of empirical fitness landscapes can be described by this model.
\end{abstract}

\setboolean{displaycopyright}{true}

\begin{document}

\maketitle
\thispagestyle{firststyle}
\marginmark
\firstpagefootnote
\correspondingauthoraffiliation{Corresponding author: Department of Physics, The Catholic University of Korea, 43 Jibong-ro, Wonmi-gu, Bucheon 14662, Republic of Korea. E-mail: spark0@catholic.ac.kr}
\vspace{-11pt}%

\lettrine[lines=2]{\color{color2}A}{} fundamental question in the theory of evolutionary adaptation concerns the distribution of mutational effect sizes and the relative roles of mutations of small vs. large effects in the adaptive process \citep{AO2005}. In his seminal 1930 monograph, Ronald Fisher devised a simple geometric model of adaptation in which an organism is described by $n$ phenotypic traits and mutations are random displacements in the trait space \citep{F1930}. Each trait has a unique optimal value and the combination of these values defines a single phenotypic fitness optimum that constitutes the target of adaptation. Because random mutations act pleiotropically on multiple traits, the probability that a given mutation brings the phenotype closer to the target decreases with increasing $n$. Fisher's analysis showed that, for large $n$, the mutational step size in units of the distance to the optimum must be smaller than $1/\sqrt{n}$ for the mutation to be beneficial with an appreciable probability. He thus concluded that the evolution of complex adaptations involving a large number of traits must rely on mutations of small effect. This conclusion was subsequently qualified by the realization that small effect mutations are likely to be lost by genetic drift, and therefore mutations of intermediate size contribute most effectively to adaptation \citep{Kimura1983,O1998,O2000E}.

During the past decade, Fisher's geometric model (FGM) has become a standard reference point for theoretical and experimental work on fundamental aspects of evolutionary adaptation \citep{Tenaillon2014}. In particular, it has been found that FGM provides a versatile and conceptually simple mechanism for the emergence of epistatic interactions between genetic mutations in their effect on fitness \citep{Martin2007,Gros2009,Blanquart2014}. For this purpose, two extensions of Fisher's original formulation of the model have been suggested. First, phenotypes are assigned an explicit fitness value, which is usually taken to be a smooth function on the trait space with a single maximum at the optimal phenotype. Second, and more importantly, mutational effects on the phenotypes are assumed to be additive. As a consequence, any deviations from additivity that arise on the level of fitness are solely due to the nonlinear mapping
from phenotype to fitness, or, in mathematical terms, due to
the curvature of the fitness function. Because the curvature is
largest around the phenotypic optimum, epistasis generally
increases upon approaching the optimal phenotype and is
weak far away from the optimum. Several recent studies have
made use of the framework of FGM to interpret experimental
results on pairwise epistastic interactions and to estimate the
parameters of the model from data \citep{Martin2007,Velenich2013,Weinreich2013,Perfeito2014,Schoustra2016}.

A particularly important form of epistatic interaction is sign
epistasis, where a given mutation is beneficial or deleterious
depending on the genetic background \citep{WWC2005}. Two types of sign epistasis are distinguished depending on whether one of the mutations affects the effect sign of
the other, but the reverse is not true [simple sign epistasis
(SSE)]; or whether the interaction is reciprocal [reciprocal
sign epistasis (RSE)]. For a pictorial representation of the
two kinds of sign epistasis, see, for example, \cite{PKWT2007}. Sign epistasis can arise in FGM either between large
effect beneficial mutations that in combination overshoot the
 fitness optimum, or between mutations of small fitness effect
 that display antagonistic pleiotropy \citep{Blanquart2014}.
 The presence of sign epistasis is a defining feature of genotypic fitness landscapes that are complex, in the sense that
 not all mutational pathways are accessible through simple
 hill climbing, and multiple genotypic fitness peaks may exist
 \citep{WWC2005,FKdVK2011,deVisser2014}. Specifically, RSE is a necessary condition for the existence of multiple fitness peaks \citep{Poelwijk2011,Crona2013}.
 Following common practice, here a genotypic fitness landscape is understood to consist of the assignment of fitness
 values to all combinations of $L$ haploid, biallelic loci that
 together constitute the $L$-dimensional genotype space. A peak
 in such a landscape is a genotype that has higher fitness than
 all its $L$ neighbors that can be reached by a single point mutation \citep{KL1987}. Note that, in contrast to
 the continuous phenotypic space on which FGM is defined,
 the space of genotypes is discrete.

\cite{Blanquart2014} showed that an ensemble of $L$-dimensional genotypic landscapes can be constructed from
 FGM by combining subsets of $L$ randomly chosen mutational
 displacements. Each sample of $L$ mutations defines another
 realization of the landscape ensemble, and the exploratory
 simulations reported by \cite{Blanquart2014} indicate a
 large variability among the realized landscapes. Nevertheless, some general trends in the properties of the genotypic
 landscapes were identified. In particular, as expected on the
 basis of the considerations outlined above, the genotypic
 landscapes are essentially additive when the focal phenotype
 representing the unmutated wild type is far away from the
 optimum and become increasingly rugged as the optimal
 phenotype is approached.

In this article we present a detailed and largely analytic study of the properties of genotypic landscapes generated under FGM. The focus is on two types of measures of landscape complexity, that is, the fraction of sign-epistatic pairs of random mutations and the number of fitness maxima in the genotypic landscape. A central motivation for our investigation is to assess the potential of FGM and related phenotypic models to explain the properties of empirical genotypic fitness landscapes of the kind that have been recently reported in the literature \citep{SSFKV2013,Weinreich2013a,deVisser2014}. The ability of nonlinear phenotype-fitness maps to explain epistatic interactions among multiple loci has been demonstrated for a virus \citep{Rokyta2011} and for an antibiotic resistance enzyme \citep{Schenk2013}, but a comparative study of several different data sets using approximate Bayesian computation (ABC) has questioned the broader applicability of phenotype-based models \citep{Blanquart2015}. It is thus important to develop a better understanding of the structure of genotypic landscapes generated by phenotypic models such as FGM.

In the next section we describe the mathematical setting and introduce the relevant model parameters: the phenotypic and genotypic dimensionalities $n$ and $L$, the distance of the focal phenotype to the optimum, and the standard deviation (SD) of mutational displacements. As in previous studies of FGM, specific scaling relations among these parameters have to be imposed to arrive at meaningful results for large $n$ and $L$. We then present analytic results for the probability of sign epistasis and the behavior of the number of fitness maxima for large $L$, both in the case of fixed phenotypic dimension $n$ and for a situation where the joint limit $n, L \to \infty$ is taken at constant ratio $\alpha = n/L$.

Similar to other probabilistic models of genotypic fitness landscapes \citep{KL1987,Weinberger1991,Evans2002,Durrett2003,Limic2004,NSK2014}, the number of maxima generally increases exponentially with $L$, and we use the exponential growth rate as a measure of genotypic complexity. We find that this quantity displays several phase transitions as a function of the parameters of FGM which separate parameter regimes characterized by qualitatively different landscape structures. Depending on the regime, the genotypic landscapes induced by FGM become more or less rugged with increasing phenotypic dimension. This indicates that the role of the number of phenotypic traits in shaping the fitness landscapes of FGM is much more subtle than has been previously appreciated, and that the sweeping designation of $n$ as (phenotypic) ``complexity'' can be misleading. Further implications of our study for the theory of adaptation and the interpretation of empirical data  will be elaborated in the \textit{Discussion}.

\section*{\zlabel{sec:Model}Model}
\subsection*{Basic properties of FGM}

In FGM, the phenotype of an organism is modeled as a set of $n$ real-valued 
traits and represented by a vector $\vec{y} = (y_1, y_2, \dots, y_n)$ in
the $n$ dimensional Cartesian space, $\vec{y} \in \mathbb{R}^n$. 
The fitness $W(\vec{y})$ is assumed to be a smooth, single-peaked function of the phenotype $\vec{y}$. 
By choosing an appropriate coordinate system, the optimum phenotype,
i.e., the combination of phenotypic traits with the highest
fitness value, can be placed at the origin in $\mathbb{R}^n$.
We also assume that the fitness $W(\vec{y})$ depends on the distance
to the optimum $|\vec{y}|$ but not on the direction of 
$\vec{y}$, which can be justified by arguments based on random matrix theory~\citep{Martin2014}. 
The uniqueness of the phenotypic optimum at the origin implies
that $W(\vec{y})$ is a decreasing function of $|\vec{y}|$. 
The form of the fitness function will be specified below when needed.  
Most of the results presented in this article are, however, independent of
the explicit shape of $W(\vec{y})$, as they rely solely on the
relative ordering of different genotypes with respect to their fitness.

When a mutation arises the phenotype of the mutant becomes
$\vec{y} + \vec{\xi}$, where $\vec{y}$ is the parental phenotype and 
the mutational vector $\vec{\xi}$ corresponds to the change of traits
due to the mutation. The key result derived by \cite{F1930} concerns the fraction
$\ProbBeneficial$ of beneficial mutations arising from a wild-type
phenotype located at distance $d$ from the optimum. Assuming that
mutational displacements have a fixed length $\vert \vec{\xi} \vert = r$ and
random directions, he showed that for $n \gg 1$
\begin{align}
	\ProbBeneficial = \frac{1}{\sqrt{2 \pi}} \int_{x}^\infty
        e^{-t^2/2} \, dt = \frac{1}{2} \text{erfc}(x/\sqrt{2}),
	\label{FisherEquation}
\end{align}
where $\mathrm{erfc}$ denotes the complementary error function and $x = r \sqrt{n}/(2d)$. Thus, for large $n$ the mutational step size has to
be much smaller than the distance to the optimum, $r \sim d/\sqrt{n}
\ll d$, for
the mutation to have a chance of increasing fitness.

As has become customary in the field, we here assume that the mutational
displacements are independent and identically distributed 
random variables drawn from an $n$-dimensional Gaussian distribution 
with zero mean. The covariance matrix can be taken to be of diagonal
form $ \sigma^2 I$, where $I$ is the $n$-dimensional identity matrix
and $\sigma^2$ is the variance of a single trait \citep{Blanquart2014}. 
In the limit $n
\to \infty$, the form of the distribution of the mutational
displacements becomes irrelevant owing to the central limit theorem (CLT),
and therefore Fisher's result of \eqref{FisherEquation} also holds in the present setting
of Gaussian mutational displacements of mean size $r = \sigma
\sqrt{n}$ \citep{Waxman2005,Ram2015}; an explicit derivation will be
provided below.
Because lengths in the phenotype space can be naturally measured in
units of $\sigma$, the parameters  
$d$ and $\sigma$ should always appear as the ratio $d / \sigma$,
as can be seen in \eqref{FisherEquation}.
Thus, without loss of generality, we can set $\sigma =1$. In the
following we denote the (scaled) wild type phenotype by $\vec{Q}$, its
distance to the optimum by 
\begin{equation}
\label{Qdef}
Q = |\vec{Q}| = \frac{d}{\sigma},
\end{equation}
and draw the displacement vectors $\vec{\xi}$ from the
$n$-dimensional Gaussian density $p(\vec{\xi}) $ with unit covariance matrix.

By normalizing phenotypic distances to the SD $\sigma$
of the mutational effect on a single trait, we are adopting a
particular pleiotropic scaling that has been referred to as the 
``Euclidean superposition model'' \citep{Wagner2008,Hermisson2008}. An
alternative choice which is closer to Fisher's original formulation but appears to have less empirical support
is the  ``total effect model,'' wherein the total length $r$ of the mutational
displacements is taken to be independent of $n$. Since $r = \sigma
\sqrt{n}$, this implies that the single trait effect size decreases
with $n$ as $\sigma \sim 1/\sqrt{n}$. As a consequence, the parameter $Q$ defined by \eqref{Qdef}
becomes $n$ dependent and increases as $\sqrt{n}$,
provided $d$ does not depend on $n$ 
\citep{O2000E}. 
The results presented below will always be given
in terms of ratios of the basic parameters of FGM, such that their
translation to the total effect model is in principle
straightforward. 
We will nevertheless explicitly point out instances
where the two settings give rise to qualitatively different
behaviors. 

\subsection*{The genotypic fitness landscape induced by FGM}

\begin{figure}
\centering
\includegraphics[width=\linewidth]{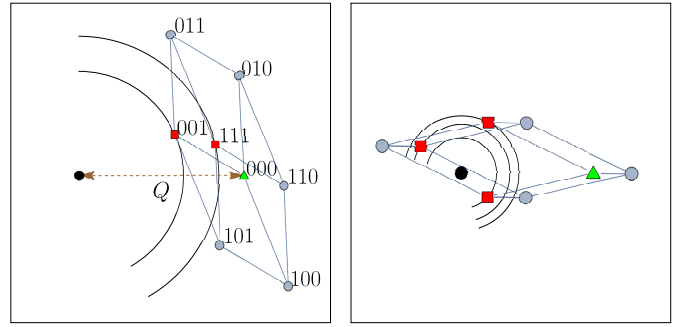}
\caption{Examples of three-dimensional genotypic fitness landscapes induced by FGM with two
  phenotypic dimensions ($L=3$ and $n=2$). The panels show the projection of the discrete
  genotype space onto the phenotype
  plane, where the phenotypic optimum is represented by a black \CIRCLE.
In the left panel, the binary sequence notation for genotypes is
indicated.  
The wild-type genotype 000, marked
by a green $\blacktriangle$, is located at distance $Q$ from the phenotypic
optimum.
The nodes represented by red $\blacksquare$'s are local fitness maxima of the
  genotypic landscapes, as can be seen from the contour lines of
  constant fitness. In the right panel the mutant phenotypes overshoot
  the optimum, whereas in the left panel they do not. 
}
\label{fig:FitnessLandscapeExamples}
\end{figure}

To study epistasis within FGM, Fisher's original definition
has to be supplemented with a rule for how the effects of multiple
mutations are combined. 
Based on earlier work~\citep{Lande1980} in quantitative genetics, 
\cite{Martin2007} introduced the assumption that mutations act additively on the level of the phenotype.
Thus the phenotype arising from two mutations $\vec{\xi_1}$, $\vec{\xi_2}$
applied to the wild-type $\vec{Q}$ is simply given by $\vec{Q} +
\vec{\xi_1} + \vec{\xi_2}$. This definition suffices to associate an
$L$-dimensional genotypic fitness landscape to any set of $L$
mutational displacements $\vec{\xi}_1, \vec{\xi}_2,\dots,\vec{\xi}_L$~\citep{Blanquart2014}.
For this purpose the haploid genotype $\tau$ is represented by a binary sequence with length $L$,
$\tau = (\tau_1,\tau_2,\ldots,\tau_L)$  with $\tau_i = 1$ ($ \tau_i =
0 $) in the presence (absence) of the $i$th mutation. 
For the wild type $\tau_i=0$ for all $i$, and in general the phenotype
vector associated with the genotype $\tau$ reads
\begin{equation}
\vec{z}(\tau) = \vec{Q} + \sum_{i=1}^{L} \tau_i \vec{\xi}_i.
\label{Eq:ztau}
\end{equation}
Two examples illustrating this genotype-phenotype map and the resulting
genotypic fitness landscapes with $L=3$ and $n=2$ are
shown in Figure~\ref{fig:FitnessLandscapeExamples}. 

As can be seen from the figure, the projection of the discrete genotype space onto the
continuous phenotype space can give rise to multiple genotypic fitness
maxima, although the phenotypic landscape is single peaked. It is the
assumption of a finite (and hence discrete) set of phenotypic mutation
vectors that distinguishes our setting from much of the earlier work on
FGM, where mutations are drawn from a continuum of alleles \citep{F1930,O1998,O2000E,AO2005}
and the probability of further improvement (as given by \eqref{FisherEquation})
vanishes only strictly at the phenotypic optimum. Remarkably, our
analysis shows that the conventional setting is not simply recovered
by taking the number of mutational vectors $L$ to infinity; rather, the
number of genotypic fitness maxima is found to increase exponentially with $L$.

Since fitness decreases monotonically with the distance to the optimum
phenotype, a natural proxy for fitness is the negative squared
magnitude of the phenotype vector
\begin{equation}
\label{Eq:distance}
- \vert \vec{z}(\tau) \vert^2 = - \vert \vec{Q} \vert^2 - 2 \sum_{i=1}^L
(\vec{Q} \cdot \vec{\xi}_i) \tau_i - \sum_{i,j = 1}^L (\vec{\xi}_i
\cdot \vec{\xi}_j) \tau_i \tau_j,
\end{equation}
where $\vec{x} \cdot \vec{y}$ denotes the scalar product between
two vectors $\vec{x}$ and $\vec{y}$. 
This quantity is thus seen to consist of a part that is additive 
across loci with coefficients given by the scalar products
$\vec{Q} \cdot \xi_i$, and a pairwise epistatic part with coefficients
$\vec{\xi}_i \cdot \vec{\xi}_j$. 

It is instructive to decompose \eqref{Eq:distance} into contributions from the
mutational displacements parallel and
perpendicular to $\vec{Q}$. Writing $\vec{\xi}_i = \xi_i^\parallel
Q^{-1} \vec{Q}
+ \vec{\xi}_i^\perp$ with $\vec{Q} \cdot
\vec{\xi}_i^\perp = 0$, \eqref{Eq:distance} can be recast into the form
\begin{equation}
\label{Eq:distance2}
- \vert \vec{z}(\tau) \vert^2 = - \left(Q + \sum_{i=1}^L
\xi_i^\parallel \tau_i \right)^2 - \sum_{i,j = 1}^L (\vec{\xi}_i^\perp
\cdot \vec{\xi}_j^\perp) \tau_i \tau_j.
\end{equation}
The first term on the right-hand side contains both additive
and epistatic contributions associated with displacements along the $\vec{Q}$ direction.
The second term is dominated by the diagonal contributions with $i=j$ and
is of order $L(n-1)$ because $ \vert \vec{\xi}_i^\perp \vert^2 = n-1$ on average.

We now show how the first term on the right-hand side of
\eqref{Eq:distance2} can be made to vanish for a range of
$Q$. For this purpose, consider the subset of phenotypic displacement
vectors for which the component $\xi_i^\parallel$ in the direction of $\vec{Q}$ is
negative. There are on average $L/2$ such
mutations, and the expected value of each component is  
\begin{equation}
\label{Eq:q0}
2 \int_{-\infty}^0 dy \, \frac{y}{\sqrt{2 \pi}} e^{-y^2/2} =
-\sqrt{\frac{2}{\pi}} \equiv -2 q_0,
\end{equation}
where the factor 2 in front of the integral arises from
conditioning on $\xi_i^\parallel < 0$. Setting $\tau_i = 1$ for $s$
out of these $L/2$ vectors and $\tau_i = 0$ for all other mutations,
the sum inside the brackets in \eqref{Eq:distance2} becomes
approximately equal to $-2q_0s$, which cancels the $Q$ term for
$s=Q/(2q_0)$. Since $s$ can be at most $L/2$ in a typical realization, 
such genotypes can be constructed with a probability approaching unity
provided $Q < q_0 L$. 

We will see below that the structure of the genotypic fitness landscapes
induced by FGM depends crucially on whether or not the phenotypes of
multiple mutants are able to closely approach the phenotypic
optimum. Assuming that the contributions from the perpendicular
displacements in \eqref{Eq:distance2} can be neglected, which will be
justified shortly, the simple argument given above shows that a close
approach to the optimum is facile when $Q < q_0L$, but becomes unlikely
when $Q \gg q_0 L$. This observation hints at a possible transition between different 
types of landscape topographies at some value of $Q$ which is
proportional to $L$. The existence and nature of this transition is
a central theme of this article. 

\subsection*{Scaling limits}
Since we are interested in describing complex organisms with large
phenotypic and genotypic dimensions, appropriate scaling relations
have to be imposed to arrive at meaningful asymptotic 
results. Three distinct scaling limits will be considered.

\begin{enumerate}[leftmargin=*]
\item Fisher's classic result (\eqref{FisherEquation}) shows
that the distance of the wild type from the phenotypic optimum has to
be increased with increasing $n$ to maintain a nonzero
fraction of beneficial mutations for $n \to \infty$. In our notation
Fisher's parameter is 
\begin{equation}
x = \frac{n}{2Q}
\label{Eq:Fisherparameter}
\end{equation}
and hence \textit{Fisher scaling} implies taking $n, Q \to \infty$ at
fixed ratio $n/Q$. We will extend Fisher's analysis by computing the
probability of sign epistasis between pairs of mutations 
for fixed $x$ and large $n$, which amounts to characterizing the shape
of genotypic fitness landscapes of size $L=2$.

\item We have argued above that the distance toward the phenotypic optimum that can be covered
by typical multiple mutations is of order $L$, and hence the limit $L
\to \infty$ is naturally accompanied by a limit $Q \to \infty$ at
fixed ratio 
\begin{equation}
\label{Eq:q}
q = \frac{Q}{L}. 
\end{equation}
From a biological point of view, one expects
that $L \gg n \gg 1$, which motivates considering the limit $L, Q \to
\infty$ at constant phenotypic dimension $n$. Under this scaling, the first term
on the right-hand side of \eqref{Eq:distance2} is of order $L^2$,
whereas the contribution from the perpendicular displacements is only
$(n-1)L$. Thus in this regime the topography of the fitness landscape
is determined mainly by the one-dimensional mutational displacements
in the $\vec{Q}$ direction, which is reflected by the fact that the
genotypic complexity is independent of $n$ to leading order and
coincides with its value for the case $n=1$, in which the
perpendicular contribution in \eqref{Eq:distance2} does not exist
(see \textit{Results}). 

\item By contrast, the perpendicular displacements play an important
role when both the phenotypic and genotypic dimensions are taken to
infinity at fixed ratio 
\begin{equation}
\label{Eq:alpha}
\alpha = \frac{n}{L}. 
\end{equation}
Combining this with the limit
$Q \to \infty$ at fixed $q = Q/L$, both terms on the right-hand side
of \eqref{Eq:distance2} are of the same order $\sim L^2$. Fisher's
parameter (\eqref{Eq:Fisherparameter}) is then also a
constant given by $x = \alpha/(2q)$. 

\end{enumerate}

\begin{table*}
	\centering
	\caption{\bf List of Mathematical Symbols}
	\begin{tableminipage}{\textwidth}
		\begin{tabularx}{\textwidth}{lX}
			\hline
			Symbols & Description \\
			\hline
			$n$ & Number of phenotypic traits, also
                        referred to as phenotypic dimension\\
			$L$ & Length of the binary genetic sequence, also referred to as genotypic dimension\\
			$\alpha$ & Ratio of phenotypic to genotypic
                        dimension ($\alpha = n/L$)\\
			$\vec{Q}$, $ Q $ & $n$-dimensional vector $\vec{Q}$ representing the wild-type phenotype and its magnitude $Q = |\vec{Q}| $\\
			$\vec{q}$, $q$ & Wild-type phenotype vector in units of $L$ ($\vec{q} = \vec{Q}/L$) and its magnitude ($q = |\vec{q}|$)\\
			$\tau$ & Genotype represented by a binary sequence  of length $L$\\
			$\tau_i$ & Binary number indicating absence (0) or presence (1) of a mutation at site $i$ ($i=1,\ldots,L$) of the genotype $\tau$\\
			$\vec{z}(\tau)$ , $z(\tau)$ & Phenotype vector
                        corresponding to the genotype $\tau$
                        (\eqref{Eq:ztau}) and its magnitude $z = \vert
                        \vec{z} \vert$\\
			$\vec{\xi}$, $\xi$ & Random phenotypic displacement vector representing a mutation and its magnitude $\xi = |\vec{\xi}|$\\
			$p(\vec{\xi})$ & Probability density of the random vector $\vec{\xi}$. In this article, the density is Gaussian with unit covariance matrix\\
			$x$ & Fisher's scaling parameter; in our notation $x = n/(2Q)$ \\
			$\mathcal{N}$ & Total number of fitness maxima in a genotypic fitness landscape averaged over all realizations of sets of $\vec{\xi}$'s\\
			$\Sigma^*$ & Genotypic complexity defined as
                        the ratio of $\ln \mathcal{N}$ to $L$ for $L
                        \to \infty$; see \eqref{Eq:Gen_Com} \\
			$q_c$  & Transition point of $q$ that separates regimes I and II\\
			$q_0$  & Half of the average mutational
                        displacement $\xi_i$  of a single trait conditioned on being
                        positive ($q_0=1/\sqrt{2\pi}$)\\
			$\rho$ & Fraction of mutations that are
                        present in a genotype $\tau$, $\rho = L^{-1} \sum_{i=1}^L \tau_i$\\
			$\rho^*$ & Mean value of $ \rho $ of a local
                        maximum, also referred to as mean
                        genotypic distance from the wild type\\
			$z^*$ & Mean value of $ z(\tau) $ of a local
                        maximum, also referred to as mean phenotypic
                        distance from the optimum\\
			\hline
		\end{tabularx}
		\label{tab:listOfNotation}
	\end{tableminipage}
\end{table*}
\subsection*{Preliminary considerations about genotypic fitness maxima}

To set the stage for the detailed investigation of the number of genotypic fitness
maxima in \textit{Results}, it is useful to develop some
intuition for the behavior of this quantity based on the elementary
properties of FGM that have been described so far. For this purpose we
consider the probability $P_\mathrm{wt}$ for the wild type to be a local fitness
maximum, which is equal to the probability that all the $L$ mutations
are deleterious. Since mutations are
statistically independent, we have
\begin{equation}
\label{Eq:PWT}
P_\mathrm{wt} = [1-\ProbBeneficial]^L = 2^{-L} \left[1+\mathrm{erf}(x/\sqrt{2})\right]^L, 
\end{equation}
where $\mathrm{erf} = 1 - \mathrm{erfc}$ is the error function. Under the (highly
questionable) assumption that this estimate can be applied to all
$2^L$ genotypes in the landscape, we arrive at the expression 
\begin{equation}
\label{Eq:NWT}
{\cal{N}}_\mathrm{wt} = 2^L P_\mathrm{wt} = \left[1+\mathrm{erf}(x/\sqrt{2})\right]^L
\end{equation}
for the expected number of genotypic fitness maxima. 

Consider first the scaling limit 2, where $x = n/(2Q) = n/(2qL) \to
0$. Expanding the error function for small arguments as
$\mathrm{erf}(y) \approx 2 y/\sqrt{\pi}$ we obtain 
\begin{equation}
\label{Eq:NWT2}
{\cal{N}}_\mathrm{wt} \approx \left[ 1 + \frac{2x}{\sqrt{2 \pi}}
\right]^L \to \exp\left( \frac{q_0 n}{q} \right)
\end{equation}
for $L \to \infty$, where $q_0 = 1/\sqrt{2 \pi}$ was defined
in \eqref{Eq:q0}. We will show below that this expression correctly
captures the asymptotic behavior for very
large $q$ but generally grossly underestimates the number of
maxima. The reason for this is that for moderate values of $q$ (in particular
for $q < q_0$), the relevant mutant phenotypes are much closer to the 
origin than the wild type, which entails a mechanism for generating a large number 
of fitness maxima that grows exponentially with $L$. 

Such an exponential dependence on $L$ is expected from \eqref{Eq:NWT}
in the scaling limit 3, where $x = \alpha/(2q)$ is a nonzero
constant and the expression in the square brackets is $>1$.
Although this general prediction is confirmed by the detailed
analysis for this case, the behavior of the number of
maxima predicted by \eqref{Eq:NWT} will again turn out to be valid
only when $q$ is very large. In particular, whereas \eqref{Eq:NWT} is
an increasing function of $\alpha$ for any $q$, we will see
below that the expected number of maxima actually decreases with
increasing phenotypic dimension (hence increasing $\alpha$) in a
substantial range of $q$. In qualitative terms, this can be attributed
to the effect of the perpendicular displacements in
\eqref{Eq:distance2}, which grows with $\alpha$ and makes it increasingly more
difficult for the mutant phenotypes to closely approach the origin.   

The observation that the number of genotypic fitness maxima grows exponentially with $L$ in most cases
motivates us to make use of the corresponding growth rate as a measure of the ruggedness of the landscape. 
We therefore define the \textit{genotypic complexity} $\Sigma^*$ through the limiting relation
\begin{align}
\Sigma^* = \lim_{L\rightarrow \infty} \frac{\ln \mathcal{N}}{L},
\label{Eq:Gen_Com}
\end{align}
where $\mathcal{N}$ is the \textit{average} number of genotypic fitness maxima and $L$ 
is the sequence length. Since the total number of binary genotypes is $2^L$, the complexity is bounded from above by $\ln 2$.
If any genotype had the same probability $P_\mathrm{max}$ of being a fitness maximum (which is in fact not the case for FGM), we could write
${\cal{N}} =  2^L P_\mathrm{max}$ and hence $P_\mathrm{max} \sim \exp[-(\ln 2 - \Sigma^\ast)L]$.

\subsection*{Data availability}
The authors state that all data necessary for confirming the conclusions presented in the article are represented fully within the article.
All numerical calculations including simulations described in this work were implemented in Mathematica and C++. When counting the number of local genotypic maxima, we checked all genotypes and counted the exact number for a randomly realized landscape, then took an average. All relevant source codes are available upon request.
\section*{Results}
\subsection*{Preliminary note}
In the following sections our results on the structure of genotypic fitness landscapes induced by FGM are stated in precise mathematical terms and the key steps
of their derivation are outlined, with some technical details relegated to the appendices.
To facilitate the navigation through the inevitable mathematical formalism, we display the definitions of the most commonly
used mathematical symbols in Table~\ref{tab:listOfNotation}.
Moreover, we provide numbered summaries at the end of each subsection which state the main results without resorting to mathematical expressions. 

\subsection*{Sign epistasis}
\subsubsection*{Random mutations:} We first study the local topography of the fitness landscape around
the wild type, focusing on the
epistasis between two random mutations with phenotypic displacements $\vec{\xi}$ and $\vec{\eta}$. 
Since fitness is determined by the magnitude of a phenotypic vector,
\textit{i.e.}, the distance of the phenotype from the origin, the
epistatic effect of the two mutations can be understood by analyzing
how the magnitudes of the four vectors
$\vec{Q}$, $\vec{Q} + \vec{\xi}$, $\vec{Q} + \vec{\eta}$ and $\vec{Q} + \vec{\xi} + \vec{\eta}$ are ordered.
To this end, we introduce the quantities 
\begin{align}
\label{Eq:R1R2R}
R_1 &\equiv \frac{1}{n} \left(|\vec{\xi} + \vec{Q}|^2 - Q^2 \right), \;\;\;
R_2 \equiv \frac{1}{n} \left( |\vec{\eta} + \vec{Q}|^2 - Q^2 \right), \text{ and }
\nonumber \\ 
R &\equiv \frac{1}{n} \left( |\vec{\xi}+\vec{\eta} + \vec{Q}|^2 - Q^2\right), 
\end{align}
where division by $n$ guarantees the existence of a finite limit for $n \to \infty$.
The sign of these quantities determines whether a mutation is beneficial or deleterious. For example,
if $R_1<0$, the mutation $\vec{\xi}$ is beneficial; if $R>0$ the two
mutations combined together confer a deleterious effect; and 
so on. We will see later that $R_{1,2}$ and $R$ are actually closely related
to the selection coefficients of the respective mutations. 

We proceed to express the different types of pairwise epistasis
defined by \cite{WWC2005} and \cite{PKWT2007} in
terms of conditions on the quantities defined in \eqref{Eq:R1R2R}. 
Without loss of generality we assume $R_1 < R_2$ and consider first
the case where both mutations are beneficial, $R_1 < R_2 < 0$. Then
magnitude epistasis (ME), the absence of sign epistasis, 
applies when the fitness of the double mutant is higher than that of each of the single
mutants, \textit{i.e.}, $R<R_1<R_2<0$. Similarly, for two deleterious mutations
the condition for ME reads $R>R_2>R_1>0$. When one mutant is
deleterious and the other beneficial, in the case of ME, the double mutant fitness has to
be intermediate between the two single mutants, which implies that 
$R_1 < R < R_2$ when $R_2 > 0 > R_1$. 

\begin{figure}
\centering
\includegraphics[width=\linewidth]{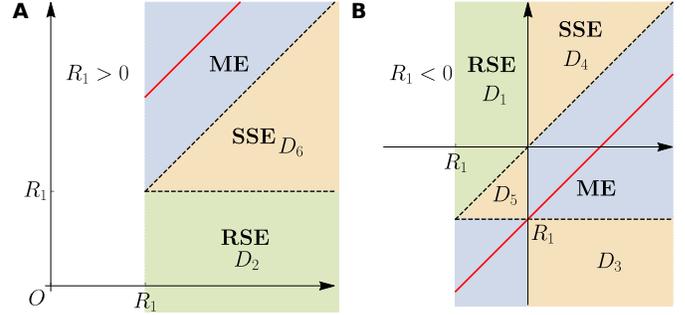}
\caption{Domains in the $(R_2, R)$ plane contributing
  to different types of epistasis: ME, SSE, and RSE.
The two panels illustrate the two cases: (A) $R_1 > 0$ and (B) $R_1 <
0$. The red solid lines indicate $R=R_1 + R_2$. The labeling of the domains $D_1, \ldots, D_6$ is used in the
derivation in \ref{appendixB:AsymptoticSignProb}. 
}
\label{fig:RegionForEachTypeOfEpistasis}
\end{figure}

The condition for RSE reads $R > R_2 > R_1$ when
both single mutants are beneficial and $R < R_1 < R_2$ when both are
deleterious, and the remaining possibility $R_1 < R < R_2$ corresponds
to SSE between two mutations of the same sign. If the two single
mutant effects are of different signs, RSE is impossible and SSE
applies when $R < R_1 < 0 < R_2$ or $R > R_2 > 0 > R_1$. 
Figure~\ref{fig:RegionForEachTypeOfEpistasis} depicts the
different categories of epistasis as regions in the $(R_2,R)$ plane. 
Note that the corresponding picture for $R_1 > R_2$ is obtained by exchanging $R_1 \leftrightarrow R_2$.

To find the probability of each epistasis, we require the joint probability density $\JointProb$.
In \ref{appendixA:DerivationOfJoinProb} it is shown that
\begin{align}
	\JointProb =& \frac{x^2 n^{1/2}}{4 \sqrt{2} \pi ^{3/2}} e^{
			-\frac{1}{8} n (-R+R_1+R_2)^2-\frac{x^2}{2} \left((R_1-1)^2+(R_2-1)^2\right)
		} \nonumber \\
		&\times  \left[1 + O\left(\frac{1}{n}\right)\right],
		\label{JointProb}
\end{align}
which can be obtained rather easily by resorting to the
CLT. The applicability of the CLT follows from
the fact that $R_{1,2}$ and $R$ are sums of a large number of independent
terms for $n \to \infty$ \citep{Waxman2005,Ram2015}. 
According to the CLT, it is sufficient to determine the first and second
cumulants of these quantities. 
Denoting averages by angular brackets,  
we find the mean  $ \Avr{R_i} = 1 $, the variance $\Avr{R_i^2} - \Avr{R_i}^2 = 1/x^2 $,
and the covariance $\Avr{R_1 R_2} - \Avr{R_1}\Avr{R_2} = 0$ ($i=1,2$).
Similarly, the corresponding quantitites evaluated for $R-R_1-R_2$ are
$ \Avr{R - R_1 - R_2} = 0 $,  $ \Avr{(R - R_1
  - R_2)^2} - \Avr{R - R_1 - R_2}^2 = 4/n $,  and $\Avr{(R-R_1-R_2)
  R_i} - \Avr{R-R_1-R_2}\Avr{R_i} = 0$ ($i=1,2$). 
With an appropriate normalization constant, this leads directly to \eqref{JointProb}.

As a first application,
we rederive Fisher's \eqref{FisherEquation}  
by integrating $\JointProb$ over the region $R_1 < 0$ for all $R_2$
and $R$, which indeed yields
\begin{align}
	\ProbBeneficial = \int_{-\infty}^{0} dR_1 \int_{-\infty}^{\infty} dR_2 \int_{-\infty}^{\infty} dR  
		\, \JointProb = \frac{1}{2}
                \text{erfc}\left(\frac{x}{\sqrt{2}}\right).
\nonumber
\end{align}
An immediate conclusion from the form of $\JointProb$ is
that it is unlikely to observe sign epistasis for large $n$, 
because $\JointProb$ becomes concentrated along the line $ R = R_1 +
R_2$ as $n$ increases. As can be seen in
Figure~\ref{fig:RegionForEachTypeOfEpistasis}, this line touches the
region of SSE in one point for $R_1 < 0$, whereas it maintains a
finite distance to the region of RSE everywhere. This indicates that
the probability of RSE decays more rapidly
with increasing $n$ than the probability of SSE. 
Moreover, one expects the latter probability to be proportional to the 
width of the region around the line $R = R_1 + R_2$, where the
joint probability in \eqref{JointProb} has appreciable weight, which
is of order $1/\sqrt{n}$.  

\begin{figure}
\centering
\includegraphics[width=\linewidth]{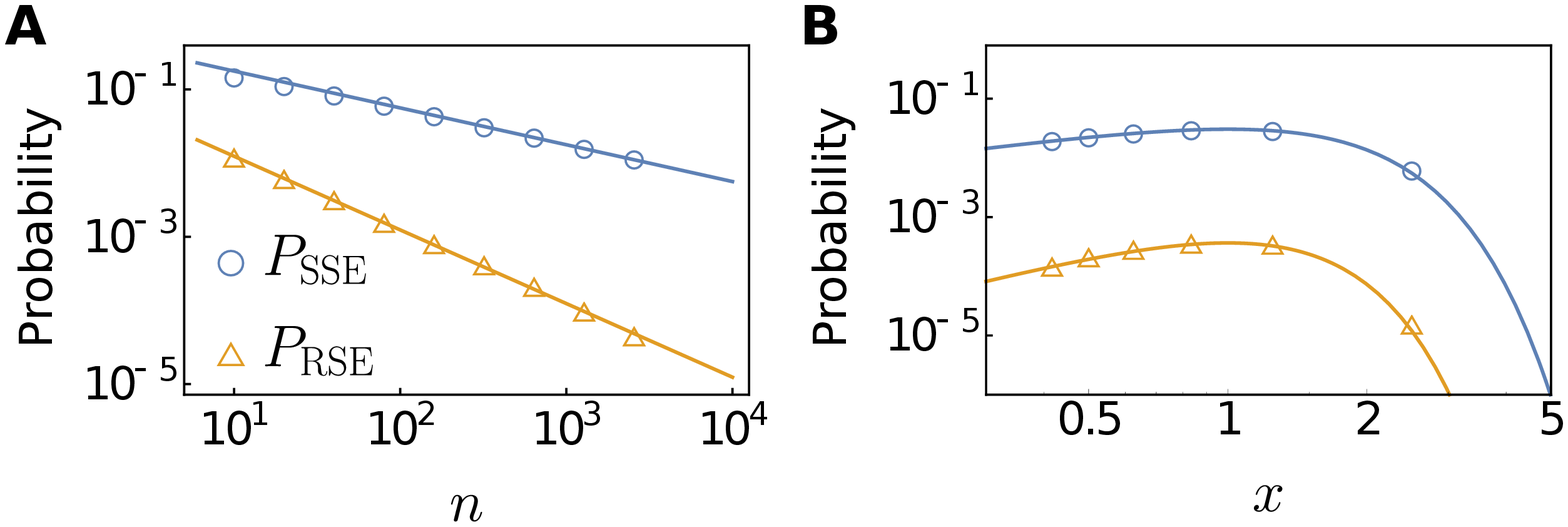}
\caption{\label{fig:CompareSimulationAlphaFixed}
Comparison of analytic results for the probability of epistasis with simulations.
Depicted are probabilities of SSE ($ \ProbSimple $) and RSE ($ \ProbReciprocal$) between two randomly chosen mutations among nearest neighbor genotypes of the wild type (A) as functions of $n$ for
fixed Fisher parameter $x=0.5$ and (B) as functions of $x$ 
for fixed phenotypic dimension $n=640$.
For each parameter set, $10^4$ randomly generated landscapes were analyzed.
The asymptotic expressions provide accurate approximations even for 
moderate $n>10$. The nonmonotonic behavior with respect to $x$ means that the probabilities are nonmonotonic functions of $Q$ for fixed $n$ and vice versa.
}
\end{figure}

To be more quantitative, we need to integrate $\JointProb$ over the
domains in Figure~\ref{fig:RegionForEachTypeOfEpistasis} corresponding to the different categories of epistasis. 
In \ref{appendixB:AsymptoticSignProb},
	we obtain the asymptotic expressions 
\begin{align}
	\ProbReciprocal =\frac{2x^2}{\pi  n} e^{-x^2}+O(n^{-3/2})
	\label{ProbReciprocal}
\end{align}
and
\begin{align}
	\ProbSimple = \frac{4 x }{\pi  \sqrt{n}}e^{-x^2/2} + O\left(n^{-1}\right)
	\label{ProbSimple}
\end{align}
for the probabilities of RSE ($\ProbReciprocal$) and SSE ($\ProbSimple$). 
Due to  the nonlinearity of the phenotype-fitness map, FGM does not
allow for strictly nonepistatic combination of fitness effects. 
The probability of ME, therefore, is
given by $\ProbMagnitude = 1 - \ProbReciprocal - \ProbSimple$.
Interestingly, the probability of sign epistasis varies
nonmonotonically with $x$. 
To confirm our analytic results, 
we compare our results with simulations in 
Figure~\ref{fig:CompareSimulationAlphaFixed},
which shows an excellent agreement. 

Similarly, we can calculate the probabilities of sign epistasis
conditioned on both mutations being beneficial, which in our setting
means $R_2 < 0$. 
The conditioning requires normalization by the unconditional
probability of two random mutations being beneficial, which is given
by the square of $P_b$ in \eqref{FisherEquation}. Hence
\begin{align}
\label{ProbBen}
	\ProbReciprocalBeneficial =\frac{2 \text{Pr}(D_1)}{P_b^2} \approx \frac{4 x ^2}{\pi  n  \,\text{erfc}\left(x/\sqrt{2}\right)^2} e^{-x^2}
\end{align}
and
\begin{align}
	\ProbSimpleBeneficial = \frac{2\text{Pr}(D_5)}{P_b^2} \approx \frac{4 x }{\pi \sqrt{n}  \,\text{erfc}\left(x/\sqrt{2}\right) }e^{-x^2/2},
\end{align}
where $\text{Pr}(D_i)$ denotes the integral of the joint probability
density over the domain $D_i$ in Figure~\ref{fig:RegionForEachTypeOfEpistasis} (see \ref{appendixB:AsymptoticSignProb}).

As anticipated from the form of \eqref{JointProb},
	the fraction of sign-epistatic pairs of mutations decreases
        with increasing phenotypic dimension $n$, and this decay
        is faster for RSE ($ \sim 1/n$) than for
        SSE ($\sim 1/\sqrt{n}$). At first glance this
        might seem to suggest that FGM has little potential for
        generating rugged genotypic fitness landscapes. However, as we
        will see below, the results obtained in this section apply
        only to the immediate
        neighborhood of the wild-type phenotype. They are modified
        qualitatively in the presence of a large number of mutations
        that are able to substantially displace the phenotype and
        allow it to approach the phenotypic optimum. 

\subsubsection*{Mutations of fixed effect size:} As a slight variation to the previous setting, one may consider the fraction of sign epistasis conditioned
	on the two single mutations to have the same selection
        strength, as recently investigated by \cite{Schoustra2016}. 
In our notation this implies that $R_1 = R_2 \equiv \tilde{R}$, and it
is easy to see that sign epistasis is always reciprocal in this case.
If the two mutations are beneficial, $\tilde{R} < 0$, and the condition
for (reciprocal) sign epistasis is $R > \tilde{R}$. The corresponding
probability is   
\begin{align}
\label{Eq:tildeP}
	\tilde{P}_\mathrm{RSE}(\tilde{R}) = \frac{\int_{\tilde{R}}^\infty \mathcal{P}(\tilde{R},\tilde{R},R) \, dR}{\int_{-\infty }^\infty \mathcal{P}(\tilde{R},\tilde{R},R) \, dR} = \frac{1}{2} \text{erfc}\left(-\frac{\sqrt{n} \tilde{R}}{2 \sqrt{2}}\right).
\end{align}
Following the same procedure for deleterious mutations ($\tilde{R}>0$)
one finds that the probability is actually symmetric around $\tilde{R} = 0$ and
hence depends only on $\vert \tilde{R} \vert$. 

To express $\tilde{P}_\mathrm{RSE}$ in terms of the selection coefficient
of the single mutations, we introduce a Gaussian phenotypic fitness
function of the form 
\begin{equation}
\label{Eq:Fitness}
W(\vec{y}) = W_0 \exp(-\lambda \vert \vec{y} \vert^2),
\end{equation}
where $\lambda > 0$ is a measure for the strength of selection. The selection coefficient of a mutation with
phenotypic effect $\vec{\xi}$ is then given by 
\begin{equation}
\label{Eq:S}
S = \ln \left[\frac{W(\vec{Q} + \vec{\xi})}{W(\vec{Q})} \right] = - \lambda \left( \vert
  \vec{Q} + \vec{\xi} \vert^2 - \vert \vec{Q} \vert^2 \right) = -
\lambda n \tilde{R}.
\end{equation}
To fix the value of $\lambda$ we note that the largest possible
selection coefficient, which is achieved for mutations that reach the phenotypic
optimum, is $S_0 = \lambda Q^2$, and hence $\tilde{R}$ is related to
the selection coefficient through
$ \tilde{R} = - (Q^2/n) (S/S_0)$.
With this substitution, the result in \eqref{Eq:tildeP} becomes
\begin{equation}
\label{Eq:tildeP2}
\tilde{P}_\mathrm{RSE}(S) = \frac{1}{2} \text{erfc}\left(\frac{n^{3/2}}{8
    \sqrt{2} x^2} \frac{\vert S \vert}{S_0}\right).
\end{equation}
The probability of sign epistasis conditioned on selection strength takes on its maximal value
$\tilde{P}_\mathrm{RSE} = 1/2$ in the neutral limit $S \to 0$ and
decreases monotonically with $\vert S \vert$. Similar to the results of
Equations~\ref{ProbReciprocal}, \ref{ProbSimple}, and \ref{ProbBen} for unconstrained mutations, it also decreases
with increasing phenotypic dimension $n$ when $S$ and $x$ are kept fixed.

\begin{figure}
\centering
\includegraphics[width=\linewidth]{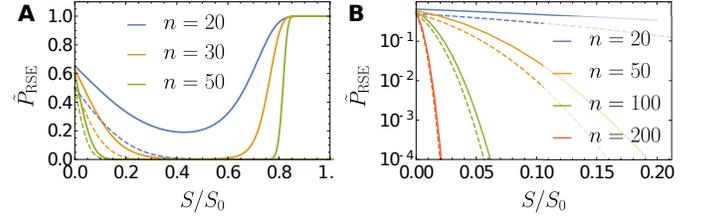}
\caption{Probability of RSE $\tilde{P}_\mathrm{RSE}$
  conditioned on the selection coefficients $S$ of the two single
  mutations to be equal and positive: (A) for the full
range of S on a linear scale and (B) for S/S0 smaller than 0.2 on a semilogarithmic
scale. Here, the fitness of a
phenotype $\vec{y}$ is assumed to be
$W(\vec{y}) = W_0 \exp(-\lambda |\vec{y}|^2)$,
where the parameter $\lambda$ is related to the maximal beneficial
selection coefficient $S_0$ through the relation $S_0 = \lambda Q^2$. 
Dashed lines depict the asymptotic expression
  \eqref{Eq:tildeP2}, and solid lines were obtained numerically using
  the Gaussian approximation for the distribution of epistasis
  developed by \cite{Schoustra2016}.
}
\label{fig:FractionSignConditioned}
\end{figure}

In a previous numerical study carried out at finite $Q$ and $n$, it was
found that $\tilde{P}_\mathrm{RSE}$ varies nonmonotonically with $S$ for the
case of beneficial mutations, and
displays a second peak at the maximum selection coefficient $S = S_0$
\citep{Schoustra2016}. The two peaks were argued to reflect the two distinct 
mechanisms giving rise to sign epistasis within FGM
\citep{Blanquart2014}. Mutations of small effect correspond to
phenotypic displacements that proceed almost perpendicularly to the direction
of the phenotypic optimum, and sign epistasis is generated through
antagonistic pleiotropy. On the other hand, for mutations of large effect, the dominant mechanism for
sign epistasis is through overshooting of the phenotypic optimum. Because
of the Fisher scaling implemented in this section with $Q, n \to \infty$ at
fixed $x = n/(2Q)$, the second class of mutations cannot be
captured by our approach and only the peak at small $S$
remains. Figure~\ref{fig:FractionSignConditioned}A shows the 
full two-peak structure for a few representative values of $n$, and
Figure~\ref{fig:FractionSignConditioned}B illustrates the convergence to the asymptotic
expression \eqref{Eq:tildeP2} for the left peak. Using the results of
\cite{Schoustra2016}, it can be shown that the right peak becomes a
step function for $n \to \infty$, displaying a discontinuous jump from $\tilde{P}_\mathrm{RSE} = 0$
to $\tilde{P}_\mathrm{RSE} = 1$ at $S/S_0 = 8/9 = 0.888\dots$.  
	
\subsubsection*{Summary 1:}
When the phenotypic dimension $n$ is large and the Fisher parameter $x$ 
is moderate, the probability of RSE decays as $1/n$, while
that of SSE decays as $1/\sqrt{n}$.
Although these probabilities decrease monotonically with $n$ at fixed $x$, they 
have a nonmonotonic behavior as a function of $x$: For small $x$ they increase with $x$
and for large $x$ they decrease with $x$ (see
Figure~\ref{fig:CompareSimulationAlphaFixed}).
Under the pleiotropic scaling adopted in this work, this implies that
the probabilities are nonmonotonic function of the wild-type distance $Q$
at fixed $n$ and vice versa. In contrast,
under the total effect model, where both the wild-type distance $Q$ and $x$ scale as $\sqrt{n}$, the probabilities decrease monotonically
and exponentially with $n$.
\subsection*{Genotypic complexity at a fixed phenotypic dimension}
In this section, we are interested in the
number of local maxima in the genotypic fitness landscape. We focus on
the expected number of maxima, which we denote by $\cal{N}$, and 
analyze how this quantity behaves in the limit of large genotypic
dimension, $L \to \infty$, when the phenotypic dimension $n$ is fixed.
For the sake of clarity, the (unique) maximum of the phenotypic
fitness landscape will be referred to as the phenotypic
\textit{optimum} throughout. 

\subsubsection*{The number of local fitness maxima:}
\begin{figure}
\centering
\includegraphics[width=\linewidth]{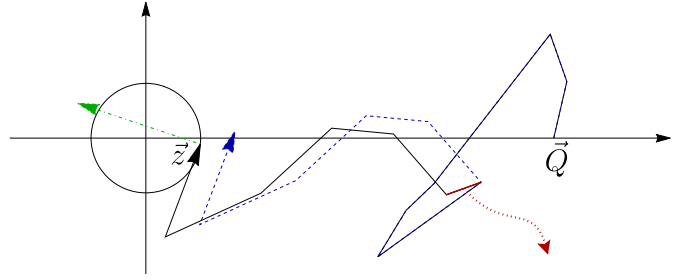}
\caption{Illustration of the condition 
for a genotype to be a local fitness maximum.
The circle encloses phenotypes that have
higher fitness than the focal phenotype $\vec{z}(\tau)$. For
$\tau$ to be a genotypic fitness maximum, both a phenotype
with a further mutation (dash-dotted green arrow) and
a phenotype without one of the mutations in $\tau$ 
(red segment and blue dotted arrows) should lie outside the circle.}
\label{fig:schematic}
\end{figure}

~Since fitness decreases monotonically with the distance to the
phenotypic optimum, a genotype $\tau$ is a local fitness maximum if
the corresponding phenotype defined by \eqref{Eq:ztau} satisfies 
\begin{align}
|\vec{z}(\tau)| < |\vec{z}(\tau) + (1 - 2 \tau_i) \vec{\xi}_i|
\label{Eq:zmaxcon}
\end{align}
for all $1 \le i \le L$. The phenotype vector appearing on the right-hand side of this inequality arises from $\vec{z}(\tau)$, either by
removing a mutation vector that is already part of the sum in \eqref{Eq:ztau}  ($\tau_i =
1$) or by adding a mutation vector that was not previously present
($\tau_i = 0$). The condition in \eqref{Eq:zmaxcon} is obviously always 
fulfilled if $\vec{z}(\tau) = 0$, that is, if the phenotype is
optimal, and we will see that in general the probability for
this condition to be satisfied is larger the more closely the
phenotype approaches the origin. A graphical illustration of the
condition in \eqref{Eq:zmaxcon} is shown in Figure~\ref{fig:schematic}. 

The ability of a phenotype $\vec{z}(\tau)$ to approach the origin
clearly depends on the number $s = \sum_{i=1}^L \tau_i$ of mutant vectors it is
composed of, and all phenotypes with the same number of mutations are
statistically equivalent. The expected number of fitness maxima
can therefore be decomposed as 
\begin{align}
	\mathcal{N} = \sum_{s=0}^{L} \binom{L}{s} \mathcal{R}_s(L),
\label{NumMaxDecomposed}
\end{align}
where $\binom{L}{s}$ is the number of possible combinations of $s$ out
of $L$ mutation vectors and $\mathcal{R}_s(L)$ is the probability that a
genotype with $s$ mutations is a fitness maximum. The latter can be written as  
\begin{align}
	\mathcal{R}_s(L) =& \int_n
d\vec{z}  
	 \left[
			\prod_{i=s+1}^{L} \int_{\mathcal{D}(-\vec{z})} d \vec{\xi}_i p(\vec{\xi}_i)
	\right]\times \nonumber\\
	&\left[
			\prod_{i=1}^{s} \int_{\mathcal{D}(\vec{z})} d \vec{\xi}_i p(\vec{\xi}_i)
		\right] 
	 \delta \left(
				\vec{z} - \vec{Q} - \sum_{i=1}^{s} \vec{\xi}_i	
	\right),
	\label{ProbLocalOptimaGivenS}
\end{align}
with
\begin{align}
\mathcal{D}(\vec{y}) \equiv \left \{ \left . \vec{\xi} \in \mathbb{R}^n \right | | \vec{\xi} -
\vec{y} | > |\vec{y}| \right \}. 
\label{DomainDefinition}
\end{align}
Here and below, $\int_n$ stands for the integral over $\mathbb{R}^n$. 

\eqref{ProbLocalOptimaGivenS} can be understood as follows. 
First, the $\delta$ function $\delta \left(\vec{z} - \vec{Q} - \sum_{i=1}^{s} \vec{\xi}_i \right)$ constrains $\vec{z}$ to be the phenotype of $\tau$ as defined in \eqref{Eq:ztau}.
Next, the integration domains of the $\vec{\xi_i}$'s reflect the condition
in \eqref{Eq:zmaxcon}. Assuming, without loss of generality, that the
$L$ genetic loci are ordered such that $\tau_i = 1$ for $i \le s$ and
$\tau_i = 0$ for $i > s$, the maximum condition for $i \le s$ requires 
$|\vec{z} | < | \vec{z} - \vec{\xi_i} | $, so the integration domain
should be $\mathcal{D}(\vec{z})$; whereas for $i>s$ the condition is
$|\vec{z}| < |\vec{z} + \vec{\xi_i}|$, corresponding to the integration domain $\mathcal{D}(-\vec{z})$.
Using the integral representation of the $\delta$ function
\begin{align}
\delta(\vec{y}) = \frac{1}{(2\pi)^n} \int_n d \vec{k} \exp\left (
i \vec{k} \cdot \vec{y} \right ),
\end{align}
we can write
\begin{align}
\mathcal{R}_s(L) = \int_n \int_n \frac{d\vec{z} d\vec{k}}{(2\pi)^n} \exp\left [
i \vec{k} \cdot \left ( \vec{z} - \vec{Q} \right ) \right ] F(\vec{k},\vec{z})^s
F(0,-\vec{z})^{L-s},
\label{Eq:Rsk}
\end{align}
where
\begin{align}
\label{Fdef}
F(\vec{k},\vec{z}) \equiv \int_{\mathcal{D}(\vec{z})} d\vec{\xi}  p(\vec{\xi})
\exp \left ( -i \vec{k} \cdot \vec{\xi}\right ).
\end{align}
It was argued on qualitative grounds in \textit{Model} that phenotypes
that approach arbitrarily close to the origin are easily generated
when the scaled wild-type distance $q$ is small, but they become
rare for large $q$. As a consequence, it turns out that the main
contribution to the integral over $\vec{z}$ in \eqref{Eq:Rsk} comes
from the region around the origin $\vec{z} = 0$ for small $q$, but
shifts to a distance $z \sim L$ along the $\vec{Q}$ direction for large $q$. To account for
this possibility, it is necessary to divide the integral domain
into two parts, $|\vec{z}| < z_0$ and $|\vec{z} | > z_0$, where
$z_0$ is an arbitrary non-zero number with $z_0/L \rightarrow 0$ as 
$L \rightarrow\infty$. Thus, we write $\mathcal{R}_s(L)$ as
\begin{align}
\label{Rsdecomp}
\mathcal{R}_s(L) = \mathcal{R}_s^<(L) + \mathcal{R}_s^>(L),
\end{align}
where
\begin{align}
\label{Rsdefs}
\mathcal{R}_s^<(L) = \int_{|\vec{z}|<z_0} d\vec{z} \int_n \frac{d\vec{k}}{(2\pi)^n} e^{
i \vec{k} \cdot ( \vec{z} - \vec{Q}  )} F(\vec{k},\vec{z})^s
F(0,-\vec{z})^{L-s}, \nonumber \\
\mathcal{R}_s^>(L) = \int_{|\vec{z}|>z_0} d\vec{z} \int_n \frac{d\vec{k}}{(2\pi)^n} e^{
i \vec{k} \cdot ( \vec{z} - \vec{Q}  )} F(\vec{k},\vec{z})^s
F(0,-\vec{z})^{L-s}, 
\end{align}
and correspondingly define 
$\mathcal{N}_<$ and $\mathcal{N}_>$ as
\begin{align}
\mathcal{N}_< = \sum_s \binom{L}{s} \mathcal{R}_s^<(L)\quad
\text{and}\quad
\mathcal{N}_> = \sum_s \binom{L}{s} \mathcal{R}_s^>(L). 
\label{Eq:Ngtlt}
\end{align}
The total number of local maxima is then 
$\mathcal{N} = \mathcal{N}_< + \mathcal{N}_>$. 

\begin{figure}
	\centering
	\includegraphics[width=\linewidth]{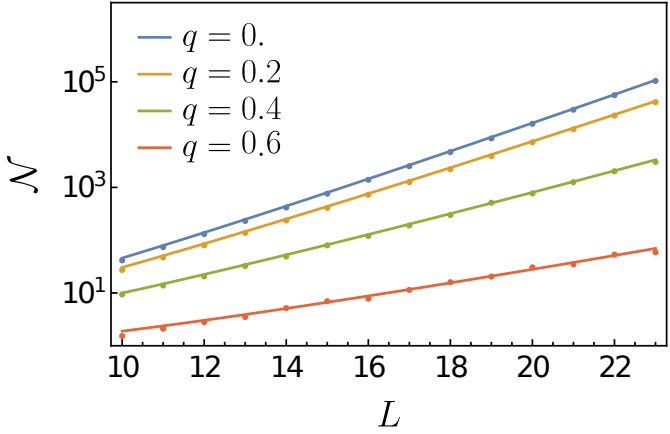}
	\caption{Plots of mean number of local maxima $\mathcal{N}$ as a function of
		the genotypic dimension $L$ for $q =0$, 0.2, 0.4, and 0.6 with $n=1$ on a semilogarithmic scale.
		Data from numerical simulations are represented as dots, and the  
		analytical prediction of \eqref{NumLocalMaximaNFixed} is shown
		as solid lines. Each dot represents the average over $10^5$ realizations of landscapes. In this parameter regime,
                $\mathcal{N}$ grows exponentially with $L$ and the
                growth rate (\textit{i.e.}, the slopes of the lines) decreases
                with increasing $q$.
	}
	\label{fig:NumericalTestComplextNFixed}
\end{figure}
\subsubsection*{Regime I:}~We first consider $\mathcal{R}_s^<(L)$. Expanding $F(\vec{k},\vec{z})$ around the
origin $\vec{z} = 0$, we show in
\ref{sec:DerivProbLocalOptima} that
\begin{align}
\mathcal{R}_s^<(L) \approx 
\frac{s^{-n/2}\exp\left [ - Q^2/(2 s) \right ] }{s \exp[-Q^2/(2s^2)] + L-s }.
	\label{RSFirstPhase}
\end{align}
For an interpretation of \eqref{RSFirstPhase} it is helpful to refer to Figure~\ref{fig:schematic}.  
Note first that the probability that 
$\vec{z} = \vec{Q} + \sum_{i=1}^s \vec{\xi}_i$ lies 
in the ball $|\vec{z}|<\zeta$ with radius $\zeta \ll 1$ is
\begin{align}
\text{Prob}(|\vec{z}|<\zeta) \approx \frac{V_{n}}{(2\pi)^{n/2}} s^{-n/2} \exp \left [ -Q^2/(2s)
\right ],
\label{Eq:probzr}
\end{align} 
where $V_n(\zeta) \sim \zeta^n$ is the volume of the ball. 
We need to estimate how small $\zeta$ has to be for $\tau$ to be a local
fitness maximum with an appreciable probability. 
Since the $s$ random vectors contributing to $\vec{z}$ are statistically equivalent, it is plausible
to assume that their average component parallel to $\vec{Q}$ is
$\xi_i^\parallel \approx -Q/s$. We further assume that the conditional
probability density $\widetilde{p}_s(\vec{\xi})$ of these vectors, conditioned on their sum $\vec{z}$ reaching
the ball around the origin, can be approximated by a Gaussian, which
consequently has the form  
\begin{align}
\widetilde p_s(\vec{\xi}) \approx \frac{1}{(2\pi)^{n/2}} \exp
\left ( -\frac{1}{2} \left |\vec{\xi} + \frac{\vec{Q}}{s} \right |^2 \right ).
\end{align}
For $\vec{z}$ to be a phenotype vector of a local maximum, all these random vectors should
lie in the region $\mathcal{D}(\vec{z})$ and the remaining
(unconstrained) $L-s$ vectors should lie in
$\mathcal{D}(-\vec{z})$. This event happens with probability 
\begin{align}
&\left [ \int_{\mathcal{D}(\vec{z})} d\vec{\xi} \widetilde p_s(\vec{\xi}) \right ]^s
\left [ \int_{\mathcal{D}(-\vec{z})} d\vec{\xi} p(\vec{\xi}) \right ]^{L-s}\nonumber \\
&\approx
\left \{1 - \frac{V_n}{(2\pi)^{n/2}} \exp[-Q^2/(2s^2)]\right \}^s \left [1 - \frac{V_n}{(2\pi)^{n/2}}\right ]^{L-s}\nonumber \\
&\approx \exp\left [- \frac{V_n}{(2\pi)^{n/2}} \left \{ s \exp[-Q^2/(2s^2)] + L-s \right \} \right ].
\end{align}
Thus, we can estimate the typical value of $\zeta$ as the solution of
\begin{align}
\frac{V_{n}(\zeta)}{ (2\pi)^{n/2}} \approx \left \{ s \exp[-Q^2/(2s^2)]+L-s\right \}^{-1},
\label{Eq:r}
\end{align}
which, combined with \eqref{Eq:probzr}, indeed gives \eqref{RSFirstPhase}.

\begin{figure}
	\centering
	\includegraphics[width=\columnwidth]{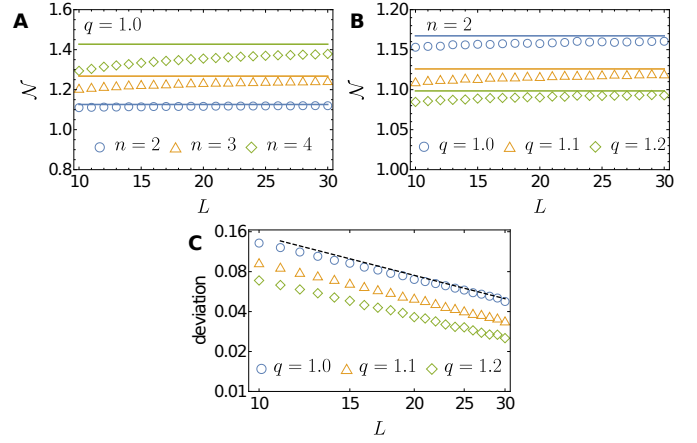}
	\caption{Comparison of simulation results (symbols) of the mean number of local 
maxima $\mathcal{N}$ with analytic approximations (lines) for $q>q_c$. 
Each symbol is the result of averaging over $2 \times 10^6$
realizations. 
(A) $\mathcal{N}$ is shown to increase with $n$ for fixed $q$. 
(B) $\mathcal{N}$ is shown to decrease with $q$ for fixed $n$. 
(C) Deviation of the analytic expression from the simulation results,
defined as $1 -
		\frac{\mathcal{N}_{\text{data}}}{\mathcal{N}_{\text{theory}}}$,
is depicted as a function of $L$ on a double logarithmic scale. 
The phenotypic dimension for this panel is $n =4$, 
where the largest deviations are observed in (A).
The deviation decreases inversely with $L$ as indicated by 
the black dashed line with slope $-1$.
	}
	\label{fig:NumericalTestComplextNFixed2}
\end{figure}

To find the asymptotic behavior of $\mathcal{N}_<$ for large $L$, we 
use Stirling's formula in \eqref{Eq:Ngtlt} and approximate the summation over $s$ by
an integral over $\rho \equiv s/L$. This yields
\begin{align}
\mathcal{N}_<  \approx 	 \int_{0}^{1} d\rho\frac{1}{ L^{n/2} \rho^{n/2}  } \frac{ e^{L \Sigma(\rho) }} {\sqrt{2 \pi  L \rho(1-\rho)  }} \frac{ 1 }{1-\rho + \rho e^{-\frac{q^2}{2 \rho ^2}}} ,
\end{align}
where the exponent $ \Sigma(\rho) $ is given by
\begin{align}
	 \Sigma(\rho) \equiv -\rho \ln\rho - (1-\rho)\ln (1-\rho) - \frac{q^2}{2\rho}.
\label{ComplexityFixedN}
\end{align}
Under the condition $L \gg 1$, the remaining integral with respect to
$\rho$ can be performed by expanding $\Sigma(\rho)$ to second order around the saddle point $\rho^*$ determined by the condition
\begin{align}
0=\left . \frac{\partial}{\partial \rho} \Sigma(\rho)\right |_{\rho = \rho^*}= \frac{q^2}{2(\rho^*)^2} - 
\ln \frac{\rho^*}{1-\rho^*}.
\label{Eq:rho*}
\end{align}
Performing the resulting Gaussian integral with respect to $\rho$ one finally obtains
\begin{align}
\mathcal{N}_<
\approx& \frac{1}{ L^{1 + n/2}  } 
\sqrt{\frac{ 1}{1+\left(1-\rho ^*\right) (q/\rho^*)^2}}  	
\frac{(\rho^*)^{-n/2}  e^{L \Sigma(\rho^*)}}{1-\rho^* + \rho^* e^{-\frac{q^2}{2 (\rho^*) ^2}}},
\label{NumLocalMaximaNFixed}
\end{align}
where $\rho^* = \rho^*(q)$ is the solution of \eqref{Eq:rho*}, which is 
the (scaled) mean number of mutations in a local maximum.
We will call $\rho^\ast$ the mean genotypic distance.
This solution is not available in closed form, but 
it can be shown that $\rho^* = (1/2) + (q^2/2) + O(q^4)$
and $\Sigma(\rho^*) = \ln 2-q^2 + O(q ^4)$ for small $q$.
Figure~\ref{fig:NumericalTestComplextNFixed} compares 
\eqref{NumLocalMaximaNFixed} with the mean number of local maxima obtained
by numerical simulations for various
$q$'s with $n=1$, to show an excellent agreement even for $L=10$.

It is obvious that $\Sigma(\rho)$ will eventually be
negative as $q$ increases for any value of $\rho$, and this must be true
also for the maximum value $\Sigma(\rho^\ast)$. Indeed, we found the threshold 
$q_c \approx 0.924~809$, above which $\Sigma(\rho^*)$ is negative. 
This signals a phase transition in the landscape properties.  
Inspection of \eqref{ComplexityFixedN} shows that the transition is
driven by a competition between the abundance of genotypes with a
certain number of mutations and their likelihood to bring
the phenotype close to the optimum. The first two terms in the
expression for $\Sigma(\rho)$ are the standard sequence entropy [see, for example, \citep{SH1997}] which
is maximal at $\rho = 1/2$ ($s = L/2$), whereas the last term represents the
statistical cost associated with ``stretching'' the phenotype toward to
origin. With increasing $q$, the genotypes contributing to the
formation of local maxima become increasingly atypical, in the sense
that they contain more than the typical fraction $\rho = 1/2$ of
mutations, and $\rho^\ast$ increases. For $q > q_c$ the cost can no longer be compensated by the
entropy term and $\Sigma(\rho^*)$ becomes negative. In
this regime $\mathcal{N}_<$ \textit{decreases} exponentially with $L$, and
therefore the total number of fitness maxima $\mathcal{N}$, which by
construction cannot be $<1$, must be dominated by the second
contribution $\mathcal{N}_>$.   

\subsubsection*{Regime II:}~We defer the detailed derivation of $\mathcal{N}_>$ to \ref{sec:DerivProbLocalOptima}
and here only report the final result obtained in the limit $L \to
\infty$, which is independent of $L$ and reads
\begin{align}
	\mathcal{N}_>
	\approx &\left[
			\frac{q-q_0}{q} \exp\left(
				{\frac{1}{q/q_0-1}}
			\right)
		\right]^{n-1}. 
	\label{NumLocalMaximaNFixedSecond}
\end{align}
This expression is valid for $q > q_0 = 1/\sqrt{2\pi} \approx
0.399$, but it dominates the contribution $\mathcal{N}_<$ for
large $L$ only when $q > q_c$. 
Figure~\ref{fig:NumericalTestComplextNFixed2} indeed shows that 
\eqref{NumLocalMaximaNFixedSecond} approximates 
the mean number of local maxima for $q>q_c$, that is, $\mathcal{N}$ converges to $\mathcal{N}_>$ 
for large $L$.
This figure also shows, as is clear by \eqref{NumLocalMaximaNFixedSecond},
that $\mathcal{N}$ is a increasing (decreasing) function of $n$ ($q$) 
for a fixed value of $q$ ($n$).
The expected number of maxima
is small in absolute terms in this regime, which can be attributed to 
the fact that the expression inside the parentheses in
\eqref{NumLocalMaximaNFixedSecond} takes the value $1.214\ldots$ at $q =
q_c$, and decreases rapidly toward unity for larger $q$.

\begin{figure}
	\centering
	\includegraphics[width=\columnwidth]{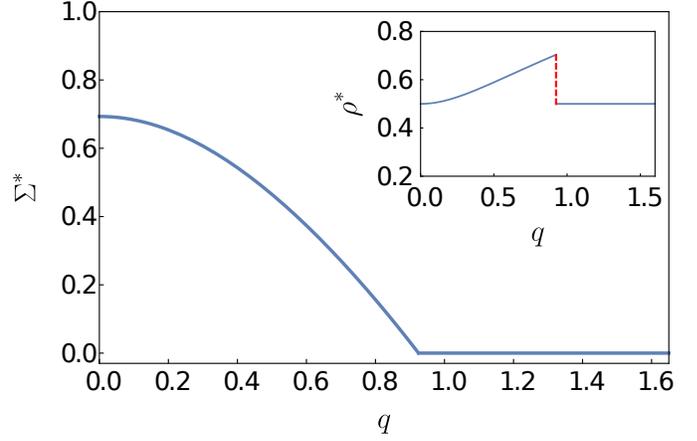}
	\caption{Plot of the genotypic complexity $\Sigma^\ast $ as a
          function of the scaled phenotypic wild-type distance
          $q$. Here the phenotypic dimension $n$ is kept finite
while taking the genotypic dimension $L$ to infinity.
		The complexity vanishes at the phase transition point
                $q = q_c\approx 0.924~809$. Inset: Plot of the mean
                genotypic distance $\rho^*$ of local maxima from the
                wild type as a function of $q$. Starting from $1/2$, $\rho^*$ increases with $q$ for $q<q_c$ and remains at $1/2$ for $q>q_c$. 
	}
	\label{fig:ComplexityNFixed}
\end{figure}
To understand the appearance of $q_0$,
we refer to \textit{Model}, where it was argued that $2 q_0 s$ 
is the maximal distance toward the origin, which can be
covered by a phenotype made up of $s$ typical mutation
vectors. Correspondingly, the analysis in \ref{sec:DerivProbLocalOptima}
shows that the main contribution to $\mathcal{R}_s^>(L)$ comes from
phenotypes located at a distance $z = 2s(q-q_0)$ from the origin,
\textit{i.e.}, at a distance $2s q_0$ from the wild type. The sum over $s$ in 
\eqref{Eq:Ngtlt} is dominated by typical genotypes with $s = L/2$, and
therefore the main contribution to $\mathcal{N}_>$ comes from
phenotypes at a distance $z = (q - q_0)L$ from the origin. 
The seeming divergence of $\mathcal{N}_>$ as $q\rightarrow q_0^+$
is an artifact of the approximation scheme, which assumes that 
the main contribution comes from the region where $z \sim
O(L)$; clearly this assumption becomes invalid when $q \to q_0^+$.
We note that for very large $q$ and large $n$,
\eqref{NumLocalMaximaNFixedSecond} reduces to the expression
$\mathcal{N}_\mathrm{wt}$ obtained in \eqref{Eq:NWT} on the basis of
Fisher's formula for the fraction of beneficial mutations from the
wild-type phenotype.

\begin{figure*}
	\centering
	\includegraphics[width=\linewidth]{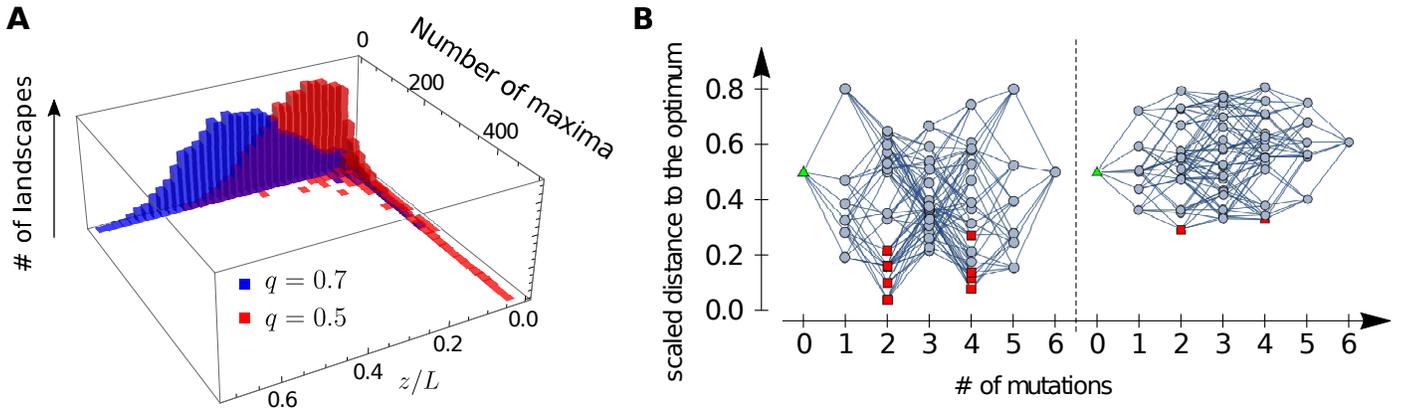}
	\caption{
		Coexistence of the two mechanisms I  and II for $q_0< q< q_c$.
		(A) Two-dimensional histogram of the number
                of fitness maxima and 
the average phenotypic distance of the maxima to the optimum within
a \textit{single} realization. Here $L=15$
and $n=2$ are used and $10^4$ different landscapes are randomly generated
for each value of $q$.
		Only a small number of realizations have a small average distance but 
these contribute an exceptionally large number of fitness peaks. 
		(B) Two examples of genotype-phenotype maps selected from
 realizations with $q=0.5$, $L=6$, and $n=2$.
		The wild type phenotype is marked by a green $\blacktriangle$
                and local fitness maxima by red $\blacksquare$'s. 
		When the phenotypes of the local fitness maxima are
                close to (far away from) the origin, 
the number of maxima is large (small), which corresponds to mechanism I (II).
	}
	\label{fig:PositionLocalMaxima}
\end{figure*}
\subsubsection*{Phase transition:}~To sum up, the leading behavior of $\mathcal{N}$ is
\begin{align}
\mathcal{N}
= \begin{cases}
\mathcal{N}_<, & q < q_c ,\\
\mathcal{N}_>, & q \ge q_c,
\end{cases}
\end{align}
with $\mathcal{N}_<$ and $\mathcal{N}_>$ given by \eqref{NumLocalMaximaNFixed} and
\eqref{NumLocalMaximaNFixedSecond}, respectively. 
Since $\mathcal{N}_<$ decreases to zero with $L$ in a 
power-law fashion at $q = q_c$, the dominant contribution at this
value is $\mathcal{N}_>$. At $q=q_c$, the mean genotypic distance $\rho^*$
jumps discontinuously from 
$\rho^*(q_c) \approx 0.7035$ to $\rho^* =
1/2$; 
and the mean phenotypic distance $z^*$, which is defined as the 
averaged magnitude of phenotype vectors for local maxima,
jumps from $z^* \approx
0$ to $z^* = (q_c - q_o) L$. The genotypic
  complexity $\Sigma^\ast$ defined in \eqref{Eq:Gen_Com} is given by 
\begin{align}
\Sigma^* = \begin{cases} \Sigma(\rho^*), & q < q_c, \\
0, & q \ge q_c,
\end{cases}
\label{Eq:S*}
\end{align}
where $\rho^*$ is the solution of \eqref{Eq:rho*}, and hence vanishes
continuously at $q=q_c$. 
These results are graphically represented in Figure~\ref{fig:ComplexityNFixed}.
Recall that the value $\Sigma^* = \ln 2$ attained at $q = 0$ is the largest possible, because 
the total number of genotypes is $2^L = \exp(L \ln 2)$. 
Remarkably, these
leading order results are independent of the phenotypic dimension. 
A dependence on $n$ emerges at the subleading order, and it
affects the number of fitness maxima in qualitatively different ways
in the two phases. For $q < q_c$,
the preexponential factor in \eqref{NumLocalMaximaNFixed} is a power
law in $L$ with exponent $1+n/2$ and hence decreases
with increasing $n$; whereas the expression in \eqref{NumLocalMaximaNFixedSecond}
describing the regime $q > q_c$ increases exponentially with $n$.

\subsubsection*{Interpretation:}

The phase transition reflects a shift between two distinct mechanisms
for generating genotypic complexity in FGM, which are
analogous to the two origins of pairwise sign epistasis that were
identified by \cite{Blanquart2014} and discussed above in \textit{Sign epistasis}.
In regime I ($q < q_c$), the mutant phenotype closely approaches the
origin and multiple fitness maxima are generated by overshooting the
phenotypic optimum. By contrast, in regime II ($q > q_c$), the
phenotypic optimum cannot be reached and the genotypic complexity
arises from the local curvature of the fitness isoclines. These two
situations are exemplified by the two panels of
Figure~\ref{fig:FitnessLandscapeExamples}. For the sake of brevity, in
the following discussion we will refer to the two mechanisms as
mechanism I and mechanism II, respectively.  

The approach to the origin in
regime I is a largely one-dimensional phenomenon governed by the components of the mutation vector
along the direction of the wild-type phenotype $\vec{Q}$, which
explains why the leading order behavior of the genotypic complexity
is independent of $n$. For $q < q_c$, the $n$-dependence of the preexponential
factor in \eqref{NumLocalMaximaNFixed} arises from the increasing difficulty of the random
walk formed by the mutational vectors to locate the origin in high
dimensions. By contrast, mechanism II operating for $q > q_c$ relies
on the existence of the transverse dimensions, which is the reason why
$\mathcal{N}_>$ in \eqref{NumLocalMaximaNFixedSecond} is an increasing function of $n$ with
$\mathcal{N}_> = 1$ for $n=1$.

When $q_0 < q < q_c$, both mechanisms seem to be present
simultaneously.
As our analysis is restricted to the average number of local maxima,
at this point we cannot decide whether both mechanisms 
appear in a single realization of the fitness landscape, 
or if one of them dominates for a given realization.
To answer this question, we generated $10^4$ fitness landscapes randomly 
for given parameter sets and identified all local maxima for each landscape.
We then determined the number of local maxima and averaged the phenotypic distance
of the local maxima to the optimum for each realization.
This mean distance will be denoted by $\tilde z$ and is itself
a random variable; it should not be confused
with the mean phenotypic distance $z^*$, which is calculated by taking
an average 
over all fitness peaks in all realizations, giving the same weight to each peak.
The results are depicted as a two-dimensional histogram in
Figure~\ref{fig:PositionLocalMaxima}A.

The figure shows that the marginal distribution of $\tilde z$ displays a pronounced peak 
around $\tilde z / L  \approx q - q_0$, which corresponds to the
behavior that is typical of mechanism I.
For most realizations,
$\tilde z/ L$ deviates significantly from zero and only a 
small number of landscapes have local maxima near $z = 0$.
However, these landscapes have many more maxima than typical
landscapes and therefore dominantly contribute to the mean number
of maxima ${\cal{N}}$. 
This shows that within a single realization the two 
mechanisms are not operative together and only a single mechanism exists. Since most realizations exhibit
mechanism II, whereas the mean number of local maxima grows
exponentially as expected for mechanism I, we conclude that mechanism
I occurs rarely but once it does, it generates a huge number of local
maxima, which compensates the low probability of occurrence. We may
thus say that both mechanisms coexist for $q_0 < q < q_c$ and $q_0$ can be regarded as the threshold
of coexistence.
Two fitness landscape realizations generated for the same value of $q$ 
located in the coexistence region that exemplify the two
mechanisms are shown in Figure~\ref{fig:PositionLocalMaxima}B.

\subsubsection*{Summary 2:} 
If the dimension $n$ of phenotypic space is
much smaller than the dimension $L$ of genotypic space, 
there exists a threshold $q_c$ of the scaled wild-type distance $q$ to the phenotypic optimum below
which the mean number $\mathcal{N}$ of local maxima in a genotypic fitness
landscape increases exponentially with $L$, and above which it saturates to a finite value.
The genotypic complexity $\Sigma^\ast$, which is defined as the exponential
growth rate of $\mathcal{N}$ with $L$, is a decreasing function of $q$ but does not depend on $n$.
On the other hand, $\mathcal{N}$ decreases with $n$ for $q<q_c$ yet
increases with $n$ for $q>q_c$. 
Figure~\ref{fig:ComplexityNFixed} depicts $\Sigma^\ast$ and
the mean genotypic distance $\rho^*$ as functions of $q$.
For $q_0 < q < q_c$, where $q_0 = 1/\sqrt{2 \pi}$, ${\cal{N}}$ is
dominated by a small fraction of landscape realizations that display
an exceptionally large number of maxima. 
If the pleiotropic scaling
is assumed to follow the total effects model,
we need to specify how the unscaled wild-type distance $d$ in
\eqref{Qdef} depends on $L$.
Assuming that $d = d_0 L$, where $d_0$ is independent of $n$ \citep{O2000E}, 
the scaled wild-type distance $q = Q/L = d_0 \sqrt{n}$
becomes an increasing function of $n$, 
and therefore the relation $q < q_c$  for regime I
is never realized when $n$ is sufficiently large.

\subsection*{Genotypic complexity in the joint limit}

In the previous subsection, we have calculated the mean number of
local fitness maxima $\mathcal{N}$
at a fixed phenotypic dimension $n$, assuming that the genotypic dimension $L$ is
much larger than $n$ ($L \gg n$).
However, in applications of FGM one often expects that both $L$ and
$n$ are large and possibly of comparable magnitude. In this case, the
results derived above can be unreliable for large $n$, as exemplified 
by the fact that the subleading correction to
\eqref{NumLocalMaximaNFixed} is of the order of $O(L^{-1/n})$ (see \ref{sec:DerivProbLocalOptima}).

\begin{figure}
	\centering
	\includegraphics[width=\linewidth]{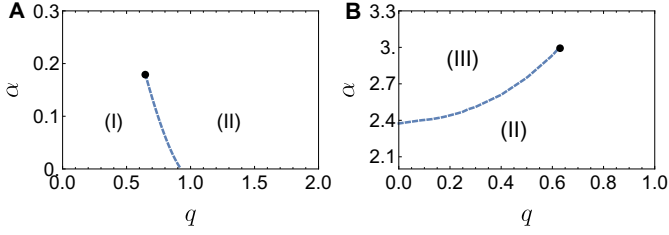}
	\caption{Phase diagrams in the parameter space $(q, \alpha)$.
Here, $q = Q/L$ is the scaled distance of the wild-type phenotype from the origin and $\alpha = n/L$ is the ratio of phenotypic dimension to genotypic dimension.
		Dashed lines are phase boundaries at which the mean genotypic and
		phenotypic distances change discontinuously. 
		(A) The phase boundary separating regimes I and II starts at $(q,\alpha) \simeq (0.925, 0)$ and continues to exist until approximately $\alpha \simeq 0.18$.
		(B) The phase boundary separating regimes II and III starts at
		$(q,\alpha) \simeq (0, 2.38)$ and continues to exist until
		approximately $q \simeq 0.62$.
	}
	\label{fig:PhaseDiagram}
\end{figure}

To obtain a reliable expression for $\mathcal{N}$ that is valid when   
both $n$ and $L$ are large, we now consider the joint limit $n, L \to
\infty$ at fixed ratio $ \alpha = n/L $. 
This will allow us to find the leading behavior of the mean number
of local maxima with a correction of order $O(1/L)$. Furthermore, we 
will clarify the role of the phenotypic dimension in the two
phases described in the previous subsection, and we will uncover a
third phase that appears at large $\alpha$ (see Figure \ref{fig:PhaseDiagram}). 

\subsubsection*{The number of local fitness maxima: }
We relegate the detailed calculation to \ref{appendixF:JoinLimit} and directly present our final expression for the mean number of local maxima,
\begin{align}
	\mathcal{N} = \mathcal{C}(a^*,b^*,g^*) e^{ L \Sigma^{\mathrm{red}}(a^*,b^*,g^*)} \left[
		1 + O\left(\frac{1}{L}\right)
	\right],
	\label{NumLocalMaximaAlpha}
\end{align}
where the function $ \Sigma^{\mathrm{red}}(a,b,g) $ in the exponent is given by
\begin{align}
	\label{ReducedComplexity}
	&\Sigma^{\mathrm{red}}(a,b,g) =
	 -\frac{\alpha}{2}\ln \left[\frac{\alpha  \left(\alpha +g\right)}{2 \left(a c(g)+b^2\right)}\right]+\frac{\alpha +2 b+g}{2}-\ln 2\nonumber\\
	&+\ln  \left\{e^{-2 c(g)} \left[ \text{erf}\left(\frac{\alpha +2 b}{\sqrt{2 a}}\right)+1\right ]+\text{erf}\left(\frac{\alpha }{\sqrt{2 a} }\right)+1\right\},
\end{align}
with $c(g) = (\alpha^2 - g^2)/(16 q^2)$. 
As before, the starred variables $a^*$, $b^*$, and $g^*$ denote the 
solution of the extremum condition
\begin{align}
	 \nabla \Sigma^{\mathrm{red}}(a,b,g) |_{(a,b,g)=(a^*,b^*,g^*)} = (0,0,0),
	 	\label{ReducedComplexityExtremumCondition}
\end{align} 
where $\nabla$ is the gradient with respect to the three
variables $(a,b,g)$. When several solutions of
\eqref{ReducedComplexityExtremumCondition} exist, the one giving the
largest value of $\Sigma^\mathrm{red}$ is chosen.  
The prefactor $ \mathcal{C}(a^*,b^*,g^*) $, which
is independent of $L$, can be determined 
from \eqref{NOptimaFinal} presented in
\ref{appendixF:JoinLimit}.
Even though the variables $(a,b,g)$ lack a direct intepretation in
terms of the original setting of FGM, 
we show in \ref{appendixFB} that $a^*$ is related to the
mean phenotypic distance $z^*$ by the equation $z^* = L \sqrt{a^*}/2 $.

An immediate consequence of \eqref{NumLocalMaximaAlpha} is that the
number of local maxima increases exponentially in $L$ for any value of
$q$ and $\alpha$ without algebraic corrections of the kind found in \eqref{NumLocalMaximaNFixed}. 
Obtaining closed-form solutions of
\eqref{ReducedComplexityExtremumCondition}, which ultimately determine the functional dependence of the complexity $\Sigma^\ast$ on $\alpha$ and $q$, seems to be a formidable task.
Instead, we resort to numerical methods by sweeping through the most interesting intervals, $q \in (0,2)$ and $\alpha \in (0,3)$.
Surprisingly, we find three independent branches of solutions that correspond to distinct phases.
To acquire a qualitative understanding of these branches, it
is instructive to first focus on the small $\alpha$ behavior, where
one expects a smooth continuation to the results of Equations \ref{NumLocalMaximaNFixed} and \ref{NumLocalMaximaNFixedSecond} as $\alpha \to 0$.

\begin{figure}
\centering
\includegraphics[width=\columnwidth]{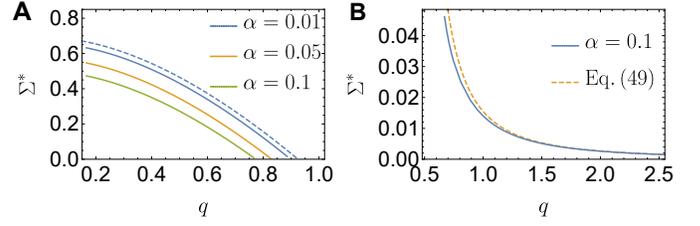}
\caption{Convergence of the complexity to the fixed $n$ case for small $\alpha$. 
(A) The solid lines depict numerical solutions of
\eqref{ReducedComplexityExtremumCondition} for values of $\alpha$
belonging to regime I.
The convergence to \eqref{NumLocalMaximaNFixed} (dashed line) is clearly seen as $\alpha \to 0$.
(B) The blue solid line depicts the numerical solution 
of \eqref{ReducedComplexityExtremumCondition} for $\alpha = 0.1$ belonging
to regime II.
Except for a slight deviation detectable when $q$ is close to $q_0$,
\eqref{SigmaII} (dashed line) remains a good approximation.
}
\label{fig:Convergence}
\end{figure}
\subsubsection*{Small $\alpha$ behavior:}
In contrast to the fixed $n$ case where two separate analyses were
carried out for the two regimes $q < q_c$ and $q > q_c$,
the present approach yields a single expression describing the genotypic complexity for
arbitrary values of $q$ and $\alpha$.
Consistently with the fixed $n$ analysis, only two out of the three
branches of solutions that were found in the numerical analysis exist
for sufficiently small $\alpha$, and they are separated by a phase
transition as shown in the phase diagram in Figure~\ref{fig:PhaseDiagram}A. 
By extrapolating the behavior of $\Sigma^*$ toward $\alpha \to 0$ as
shown in Figure~\ref{fig:Convergence}, we are able to identify the correct
counterparts for each of the two previously found regimes.

\begin{figure*}
	\centering
	\includegraphics[width=\textwidth]{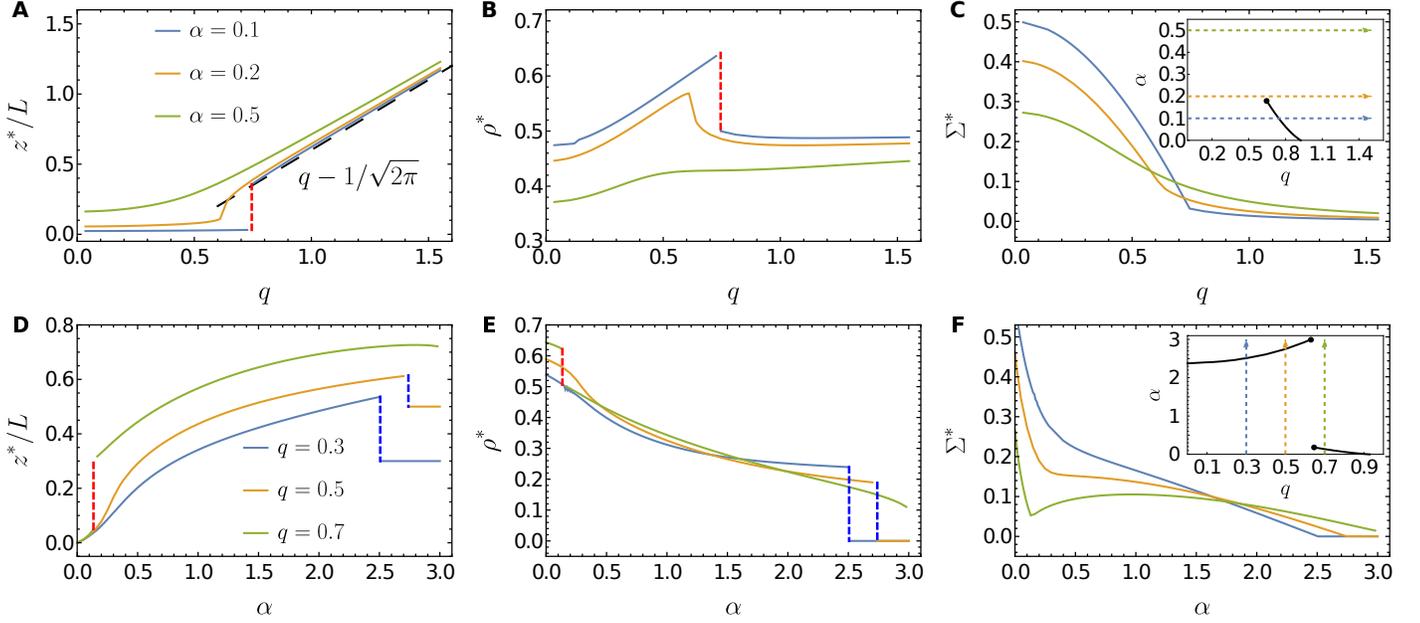}
	\caption{Plots of scaled mean phenotypic distance $z^\ast/L$ (left column), 
mean genotypic distance $\rho^\ast$ (middle column), and 
genotypic complexity $\Sigma^*$ (right column) against $q$ for fixed $\alpha$ (top row)
and against $\alpha$ for fixed $q$ (bottom row).
The curves in the top (bottom) panels are drawn along the arrows in the inset of (C) and F.
	Top row (A, B, and C): When  $\alpha$ is small, the landscape behaves
similar to the fixed $n$ case which effectively corresponds to $\alpha =0$.
In this case $z^*$ and $\rho^*$ for large $q$ are well approximated by $q - q_0$ and
$1/2$, respectively.
	As $\alpha$ increases beyond the transition line, the first-order transition visualized by the red dashed lines 
		disappears and all quantities change smoothly with $q$.
	Bottom row (D, E, and F): 
	As $\alpha$ increases for small $q$, another phase
        transition with discontinuities in $z^\ast$ and $\rho^\ast$
        (blue dashed lines) signals the appearance of regime III. The genotypic maxima in regime III are located very close to 
	the wild-type position, $z^\ast/L \simeq q$ and $\rho^\ast \simeq 0$. 
	This transition ceases to exist when $q$ exceeds approximately 0.62.
	Note that the dependence of $\Sigma^\ast$ on $\alpha$ is
        nonmonotonic for $q=0.7$ (F).  
	}
	\label{fig:ComplexityAlphaLimit}
\end{figure*}

The extrapolation is straightforward in regime II, 
where the replacement $n  \to  \alpha L$ in
\eqref{NumLocalMaximaNFixedSecond} yields an exponential
dependence of $\mathcal{N}$ on $L$ with the growth rate
\begin{align}
\Sigma^{*\mathrm{(II)}}_\mathrm{approx} = \alpha \ln \left\{
	\frac{q-q_0}{q} \exp\left[
		(q/q_0-1)^{-1}
	\right]	
\right\}.
\label{SigmaII}
\end{align}
This crude approximation turns out to be remarkably accurate even at $\alpha = 0.1$, as illustrated in Figure \ref{fig:Convergence}B.
By contrast, in regime I the naive replacement of $n$ by $\alpha L$ in
\eqref{NumLocalMaximaNFixed} yields an expression that vanishes faster
than exponential in $L$, as $\exp[ - (\alpha/2) L \ln L]$. 
This reflects the fact that the mean phenotypic distance $z^*$ 
moves away from the origin for any $\alpha > 0$
and hence the complexity cannot be derived only by inspecting \eqref{ProbLocalOptimaGivenS} around $z=0$ (see Figure \ref{fig:ComplexityAlphaLimit}, A and D).
At the same time, the mean genotypic distance $\rho^\ast$
decreases with increasing $\alpha$ and
eventually falls below the value $\rho^\ast = 1/2$ favored by the sequence
entropy (Figure \ref{fig:ComplexityAlphaLimit}, B and E).

Both trends can be attributed to the increasing role of the perpendicular
mutational displacements that make up the second term on the right-hand
side of \eqref{Eq:distance2}. Under the scaling of the joint limit, this term is of order $\rho L (n-1)
\approx \rho \alpha L^2$ and hence comparable to the first term
originating from the parallel displacements. The perpendicular
displacements always increase the phenotypic distance to the origin, and they are
present even when $q=0$. The additional cost to reduce the 
perpendicular contribution results in a smaller value of $\Sigma^*$
compared to the case of fixed $n$. Moreover, whereas the parallel
contribution is minimized (for $ q > q_0$) by making $\rho$ as large
as possible, the reduction of the perpendicular displacements requires small $\rho$. 

In the fixed $n$ analysis, the number of fitness maxima was found to 
decrease (increase) with $n$ in regime I (II)
and this tendency is recovered from the joint-limit case
when $\alpha$ is not too large (Figure \ref{fig:ComplexityAlphaLimit}C). 
Because of these opposing trends of $\Sigma^*$ in the two regimes, the
location of the phase transition separating them is expected to
decrease with increasing $\alpha$, as can be seen in Figure \ref{fig:PhaseDiagram}A. 
If one ignores the contribution from the perpendicular displacements,
the phenotypic position of the fitness maxima is expected to
jump from $z^\ast = 0$ to $z^\ast = q_c-q_0$ at the transition, and
thus the jump size should decrease as $q_c$ decreases.
This observation suggests that the two branches should merge into one
when $q_c$ reaches $q_0$.
With the additional contribution of perpendicular dimensions, we
numerically found that this critical end point 
at which the phases I and II merge occurs even earlier, at $\alpha
\simeq 0.18$ and $q \simeq 0.62 > q_0$
(Figure \ref{fig:PhaseDiagram}). For $\alpha > 0.18$, 
$\rho^*$ does not show any discontinuity for any $q$ as long as
the parameters are in regime II.

\subsubsection*{Large $\alpha$ behavior and regime III:}
To develop some intuition about the FGM fitness landscape in
the regime where $\alpha = n/L \gg 1$, we revisit the results
obtained in \textit{Sign epistasis}, 
where pairs of mutations were considered. Two
conclusions can be drawn about the typical shape of these small
genotypic landscapes (of size $L=2$) in the limit $n \to
\infty$. First, the probability that the wild type
is a genotypic maximum tends to unity according to \eqref{Eq:PWT}. Second,
the joint distribution given in \eqref{JointProb} enforces additivity of
mutational effects for large $n$, and correspondingly the
probability for sign epistasis vanishes. Thus for large $n$ the
two-dimensional genotypic landscape becomes smooth with a single
maximum located at the wild type. 
Assuming that this picture holds more generally whenever the limit $n
\to \infty$ is taken at finite $L$, we expect the following asymptotic
behaviors of the quantifiers of genotypic complexity for large
$\alpha$: (i) $\mathcal{N} \to 1$, $\Sigma^* \to 0$ (unique genotypic
optimum); and (ii) $z^\ast/L \to q$, $\rho^\ast \to 0$ (location of the
maximum at the wild-type phenotype and genotype).

This expectation is largely borne out by the numerical results shown
in the bottom panels of Figure~\ref{fig:ComplexityAlphaLimit}. 
However, depending on the value of
$q$, the approach to the limit of a smooth landscape can be either continuous (for large $q$) or display characteristic jumps as indicated
by the blue dashed lines in Figure \ref{fig:ComplexityAlphaLimit}, D
and E. These jumps as well as the discontinuity in the
slope of $\Sigma^*$ as a function of $\alpha$ in Figure
\ref{fig:ComplexityAlphaLimit}F are hallmarks of the phase
transition to the new regime III, which is represented by the dashed
line in Figure \ref{fig:PhaseDiagram}B. 

Fortunately, the solution of
\eqref{ReducedComplexityExtremumCondition} describing the new phase
can be obtained analytically from \eqref{ReducedComplexity} or
\eqref{ReducedComplexityRho} in \ref{appendix:AFB} as a series expansion.
The derivation presented in \ref{appendix:ThirdSaddlePoint} yields
\begin{align}
a^* &= 4q^2 - \left[
\frac{16 \sqrt{\frac{2}{\pi }} q^3}{\alpha ^2} + O(q^4/\alpha^3)
\right] \epsilon + O(\epsilon^2), \nonumber \\
b^* &= - \alpha +\frac{\alpha  \epsilon}{\sqrt{2 \pi } q} + O(\epsilon^2), \nonumber \\
g^* &= \alpha + O(\epsilon^2), \quad \textrm{and} \quad
\rho^* = \frac{\sqrt{\frac{2}{\pi }} q \epsilon }{\alpha } +  O(\epsilon^2),
\label{SaddlePointThird}
\end{align}
where the expansion parameter $\epsilon = e^{-\alpha ^2/(8
    q^2)}$ decays rapidly with increasing $\alpha/q$. 
The corresponding genotypic complexity can also be evaluated in a series expansion,
\begin{align}
	\Sigma^{\textrm{(III)}}(a^*, b^*, c^*) =  \frac{\alpha  \epsilon ^2}{4 \pi  q^2} + O(\epsilon^3),
\end{align}
which shows that $\Sigma^*$ is positive
but vanishingly small in this regime.
We note that using \eqref{RelationAandZ}, the expression for $a^*$ in
\eqref{SaddlePointThird} amounts to 
\begin{align}
\frac{z^*}{L} \simeq q -\frac{2 \sqrt{\frac{2}{\pi }} q^2 }{\alpha ^2}
  \epsilon, 
\label{zstarThird}
\end{align}
implying that the small number of local maxima that exist in this
phase are located very close to the wild-type phenotype.

To first order in $\epsilon$, the results for $\rho^\ast$ and $z^\ast$
in Equations \ref{SaddlePointThird} and \ref{zstarThird}
can be easily derived from the idea that mutational effects become approximately additive
for large $\alpha$, thus providing further support for
this assumption. If mutational effects are strictly additive, the probability
for a genotype contaning $s$ mutations to be a local fitness maximum
is given by 
\begin{align}
\label{Rsadditive}
\mathcal{R}_s^\mathrm{add} = P_b^s (1-P_b)^{L-s},
\end{align}
where $P_b$ is the probability for a mutation to be
beneficial. \eqref{Rsadditive} expresses the condition that reverting
any one of the $s$ mutations contained in the genotype as well as adding one of
the unused $L-s$ mutations should lower the fitness. 
Using Fisher's \eqref{FisherEquation}, the probability
for a beneficial mutation is $P_b \approx \sqrt{(2/\pi)} (q/\alpha)
  \epsilon$ for large $\alpha$. Thus to linear order in
$\epsilon$ or $P_b$, the expected number of mutations contributing to
such a genotype is  $ L P_b = L \rho^* = L \sqrt{(2/\pi)} (q/\alpha) \epsilon$, which is consistent with \eqref{SaddlePointThird}.

The phenotypic location of a local maximum deviates from $\vec{Q}$ 
in those rare instances where one of the mutations from the
wild type is beneficial, which happens with probability $P_b$.  
To estimate the corresponding shift in $z^\ast$, we refer to the
results of subsection \textit{Sign epistasis}, where it was shown that
the squared phenotypic displacement $R_1$ defined in \eqref{Eq:R1R2R}
has a Gaussian distribution with mean 1 and variance $1/x^2 =
4q^2/\alpha^2$ for large $n$. Using this, it is straightforward to
show that the expected value of $R_1$ conditioned on the
mutation to be beneficial ($R_1 < 0$) is $\bar{R}_1 = -4
  q^2/\alpha ^2$ to leading order. Multiplying this by the expected
number of mutations $L P_b$ we obtain the relation
\begin{align} 
L \rho^* \bar{R}_1 \approx \frac{(z^*)^2 - Q^2}{n},
\end{align}
which yields the same leading behavior for $z^*/L$ as in
\eqref{zstarThird}. 

\begin{figure}
	\centering
	\includegraphics[width=\columnwidth]{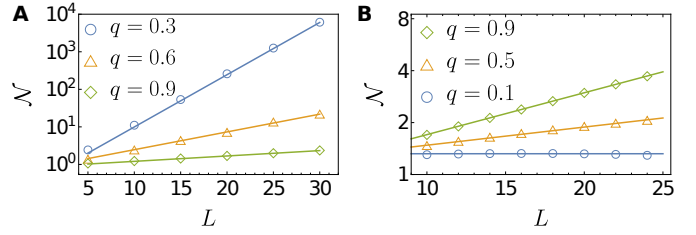}
	\caption{Semilogarithmic plots of the mean number of local maxima $\mathcal{N}$
vs. the genotypic dimension $L$ for (A) $\alpha = 0.2$ and
(B) $\alpha = 2.5$ and for various values of $q$.
Each symbol represents the average over $10^5$ randomly generated landscapes, 
and lines depict the analytic approximation of \eqref{NOptimaFinal}.
The approximation is good even for moderate $L$.
	}
	\label{fig:NumericalTest}
\end{figure}

As previously observed for the transition between regimes I and II,
the phase boundary separating regimes II and III terminates at a point
where the two solutions defining the regimes merge (Figure
\ref{fig:PhaseDiagram}B). Beyond this point the jumps in $z^\ast$ and $\rho^\ast$
seen in Figure \ref{fig:ComplexityAlphaLimit}, D and E, disappear and
all quantities approach smoothly to their asymptotic values. 
A surprising feature of the large $\alpha$ behavior that persists
also for larger $q$ is that the complexity becomes an \textit{increasing}
function of $q$ when $\alpha > 1.7$ (Figure
\ref{fig:ComplexityAlphaLimit}F). In Figure \ref{fig:NumericalTest}
we verify this behavior using direct simulations of FGM. These
simulations also show that the predictions based on \eqref{ReducedComplexity} are already 
remarkably accurate for moderate values of $L$ and $n$. 
\subsubsection*{Summary 3:} When the dimension $n$ of the phenotypic trait space 
and the dimension $L$ of the genotypic space are large and comparable,
the genotypic complexity $\Sigma^\ast$ is always nonzero and depends on the 
ratios $\alpha = n/L$ and $q = Q/L$.
There are three regimes where the behavior of the genotypic complexity 
and the mean genotypic distance $\rho^*$ (the average number
of mutations in a local maximum divided by $L$)
are qualitatively different.
In regime I, which is roughly characterized by small $q$ and small $\alpha$,
there are many local maxima in the region located far away from the
wild type but close to the phenotypic optimum,
and the fitness landscape is quite rugged.
In regime II, which is roughly characterized by large $q$ and small 
$\alpha$, there is an appreciable number of local maxima, though smaller than
in regime I, and typically half of the $L$ mutations contribute to the
corresponding genotypes.
In regime III, which is roughly characterized by large $\alpha$, the genetic complexity is very small, though nonzero. 
Also $\rho^*$ is close to zero, which means that the wild type has a high probability
to be the global fitness maximum. An overview of the three regimes is found in Table~\ref{tab:Sum}.


\begin{table}
	\centering
	\caption{\bf Characteristics of the three regimes in the joint limit}
	\begin{tableminipage}{\columnwidth}
		\begin{tabularx}{\columnwidth}{cXccX}
			\hline
			Regime  & Condition & $\Sigma^*$ &$\rho^\ast$ &  landscape \\
			\hline
			I	& $q \ll 1$, $\alpha \ll 1$ & $>0$ & $>1/2$& rugged\\
			II	& $q \gg 1$, $\alpha \ll 1$ & $\simeq 0$ &$\approx 1/2$ & intermediate\\
			III	& $\alpha \gg 1$ & $\simeq 0$ & $\simeq 0$&  almost smooth\\
			\hline
		\end{tabularx}
		\label{tab:Sum}
	\end{tableminipage}
\end{table}
\section*{Discussion}

FGM provides a simple yet generic scenario for
the emergence of complex epistatic interactions from a nonlinear
mapping of an additive, multidimensional phenotype onto fitness. Its
role in the theory of adaptation may be aptly described as that of a 
``proof-of-concept model'' \citep{Servidio2014}, and as such it is
widely used in fundamental theoretical studies
\citep{Blanquart2014,Chevin2010,Fraisse2016,Gros2009,Martin2014,Moura2016} as well as for the
parameterization and interpretation of empirical data \citep{Bank2014,Blanquart2015,Martin2007,Perfeito2014,Schoustra2016,Velenich2013,Weinreich2013}. Rather than
tracing the mutational effects and their interactions to the
underlying molecular basis, the model aims at identifying robust
features of the adaptive process that can be expected to be shared by large classes of organisms. 

To give an example of such a feature that is of central importance in
the present context, it was pointed out by
\cite{Blanquart2014} that pairwise sign epistasis is generated in FGM through two
distinct mechanisms. In one case the mutational
displacements overshoot the phenotypic optimum, whereas in the other
case the displacements are directed approximately perpendicular to the
direction of the optimum, and sign epistasis arises because the fitness
isoclines are curved. The first mechanism is obviously also operative in
a one-dimensional phenotype space, but in the second case \citep[termed
antagonistic pleiotropy by][]{Blanquart2014} at least two
phenotypic dimensions are required. 
Interestingly, both mechanisms
have been invoked in empirical studies where a nonlinear
phenotype-fitness map was used to model epistatic interactions
between multiple mutations. In one study, \cite{Rokyta2011} explained the pairwise epistatic interactions
between nine beneficial mutations in the single-stranded DNA bacteriophage
ID11 by assuming that fitness is a
single-peaked nonlinear function of a one-dimensional additive
phenotype. In the second study the genotypic
fitness landscapes based on all combinations of two groups of four antibiotic
resistance mutations in the enzyme $\beta$-lactamase were parameterized
by a nonlinear function mapping a two-dimensional phenotype to
resistance \citep{Schenk2013}. The fitted function was in fact monotonic
and did not possess a phenotypic optimum, which makes it clear that
the epistatic interactions arose solely from antagonistic pleiotropy
in this case. 

In this work we have shown that the two mechanisms described by
\cite{Blanquart2014} lead to distinct regimes or phases in the
parameter space of FGM, where the genotypic fitness landscapes
display qualitatively different properties (Figure \ref{fig:PhaseDiagram}A). When the phenotypic dimension $n$
is much smaller than the genotypic dimension $L$, the two regimes are
separated by a sharp phase transition where the average number and location of
genotypic fitness maxima changes abruptly as the distance $q$ of the
wild-type phenotype from the optimum is varied. In regime I ($q < q_c$), the
phenotypic optimum is reachable at least by some combinations of
mutational displacements. Overshooting of the optimum is therefore
possible and sign epistasis is strong, leading to
rugged genotypic landscapes with a large number of local fitness maxima 
that grows exponentially with $L$. By contrast, in regime II ($q >
q_c$), only antagonistic pleiotropy is operative and the number of
fitness maxima is much smaller. More precisely, for finite $n$ the
number tends to a finite limit for $L \to \infty$, but the limiting
value is an exponentially growing function of $n$. 

An important consequence of our results is that the dependence of the
fitness landscape ruggedness on the phenotypic dimension $n$ is remarkably
complicated. For $n \ll L$, landscapes become less rugged with increasing
$n$ in regime I ($q < q_c$) but display increasing ruggedness in
regime II ($q > q_c$). When $n \gg L$ the ruggedness decreases with $n$ for all
$q$ and the landscapes become approximately additive (regime
III). In particular, the probability of sign epistasis vanishes
algebraically with $n$ in this regime. Thus $n$ cannot in general be
regarded as a measure of ``phenotypic complexity,'' as a
larger value of $n$ does not imply that the corresponding fitness
landscape is more complex.  

\begin{figure}
\centering
\includegraphics[width=\linewidth]{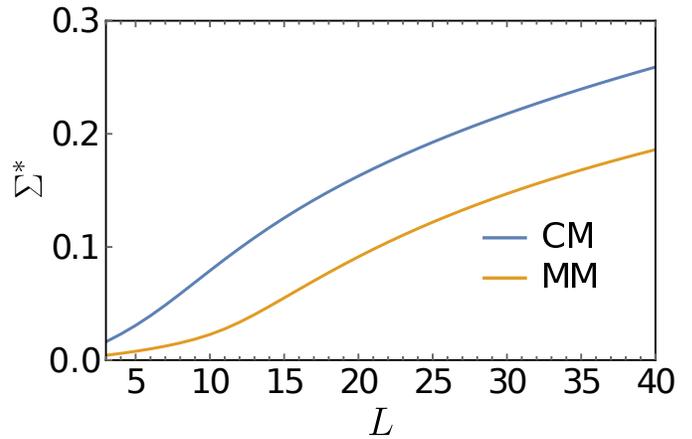}
\caption{The logarithm of the number of local fitness maxima divided
  by the number of loci $L$ is shown as a function of $L$
  for FGM with the parameter values $n=19.3, Q = 6.89$ and $n= 34.8,
  Q= 9.81$ obtained by \cite{Schoustra2016} for the fungus
  \textit{A. nidulans} growing in complete (CM) and minimal medium (MM),
  respectively. For the evaluation of $\mathcal{N}$
  \eqref{ReducedComplexity} was used.
}
\label{fig:Nidulans}
\end{figure}

This observation is relevant for the interpretation of experiments
where the parameters of FGM are estimated from data. In recent work,
FGM was used to analyze data on pairwise epistasis between beneficial mutations
in the filamentous fungus \textit{Aspergillus nidulans} growing in two
different media \citep{Schoustra2016}. The estimates obtained for the
phenotypic dimension and the distance of the wild-type phenotype from
the optimum were $n=19.3, Q = 6.89$ in complete medium and $n= 34.8,
Q= 9.81$ in minimal medium, which, surprisingly, may seem to suggest a higher
phenotypic complexity in the minimal medium. Using the results derived in this
article, we can translate the estimated parameter values into the
average number of maxima that a genotypic fitness landscape of a given
dimension $L$ would have. As can be seen in Figure \ref{fig:Nidulans},
with respect to this measure the fitness landscape 
of the fungus growing in complete medium is actually more
rugged. This is consistent with experiments using \textit{Escherichia
  coli}, which found a greater heterogeneity of fitness trajectories
in complete medium \citep{RHHD2008}, and indicates that the complete medium allows for
a greater diversity of paths to adaptation than the minimal medium. 

We hope that the results presented here will promote the use of 
FGM as part of the toolbox of probabilistic models that are
currently available for the analysis of empirical fitness landscapes
\citep{Bank2016,Blanquart2015,deVisser2014,Hayashi2006,NSK2014,SSFKV2013}. Compared
to purely genotype-based models such as the NK and rough-Mount-Fuji
(RMF) models, FGM is arguably more realistic in that it introduces an
explicit phenotypic layer mediating between genotypes and fitness
\citep{Martin2014}. Somewhat similarly to the RMF model, the fitness landscapes of FGM
are anisotropic and display a systematic change of properties as a
function of the distance to the optimal phenotype (FGM) or the
reference sequence (RMF), respectively \citep{NSK2014}. The idea 
that fitness landscape ruggedness increases systematically and
possibly abruptly when approaching the optimum has been proposed
previously in the context of \textit{in vitro} evolution of proteins
\citep{Hayashi2006}. If this is indeed a generic pattern, it may
have broader implications.
For example, \cite{dVPK2009} showed that the evolutionary benefits
of recombination are severely limited by the presence of multiple
peaks. If such peaks are rarely encountered far away from the optimum, the benefits
of recombination would be most pronounced for particularly maladapted populations. 
 
A recent investigation of 26 published empirical fitness landscapes using 
ABC concluded that FGM could
account for the full structure of the landscapes only in a minority
of cases \citep{Blanquart2015}. One of the features of the 
empirical landscapes that prevented a close fit to FGM was the
occurrence of sign epistasis far away from the phenotypic optimum.
Our analysis confirms that this is an unlikely event in FGM, 
and precisely quantifies the corresponding probability through
\eqref{ProbReciprocal} and \eqref{ProbSimple}. \cite{Blanquart2015} also found that the
phenotypic dimension is particularly difficult to infer from
realizations of genotypic fitness landscapes,  which matches our
observation that the structure of the landscape depends only weakly on
$n$ when $n \ll L$. We expect that our results will help to further
clarify which features of an empirical fitness landscape make it
more or less amenable to a phenotypic description in terms of FGM or
some generalization thereof. 

We conclude by mentioning some open questions that should be addressed
in future theoretical work on FGM. First, a significant limitation of
our results lies in their restriction to the average 
number of local fitness maxima. The number of maxima
induced by a given realization of mutational displacements is a
random variable, and unless the distribution of this variable is well
concentrated, the average value may not reflect the typical behavior. 
The large fluctuations between different realizations of fitness
landscapes generated by FGM were noticed already by
\cite{Blanquart2014} on the basis of small-scale simulations,
and they clearly contribute to the difficulty of inferring the
parameters of FGM from individual realizations that was reported by
\cite{Blanquart2015}.  In light of our analysis, this pronounced heterogeneity can be attributed to the
existence of multiple phases in the model, and it is exemplified by
the simulation results in Figure \ref{fig:PositionLocalMaxima}. To
quantitatively characterize the fluctuations between different
realizations, a better understanding of the distribution of the number 
of fitness maxima and its higher moments is required. 

Second, the consequences of relaxing some of the
assumptions underlying the formulation of FGM used in this
work should be explored. The level of pleiotropy can be reduced by restricting the
effects of mutational displacements to a subset of traits
\citep{Chevin2010,Moura2016}, and it would be interesting to see how
this affects the ruggedness of the fitness landscape. However, the most
critical and empirically poorly motivated assumption of FGM is clearly
the absence of epistatic interactions on the level of phenotypes. It
would therefore be important to understand how robust the results
presented here are with respect to some level of phenotypic epistasis,
which should ideally arise from a realistic model of phenotypic networks 
\citep{Martin2014}. 

Third, a natural extension of the present study is to consider multiallelic 
genetic sequences.  
An immediate generalization keeping the additivity of mutational effects on the 
level phenotypes is to consider the following genotype-phenotype map:
\begin{align}
\vec{z}(\tau) = \vec{Q} + \sum_{i=1}^L \sum_{k=1}^A \tau_{ik} \vec{\xi}_{ik},
\end{align}
where $A$ is the size of the alphabet from which the sequence elements
are drawn (\textit{e.g.}, $A=4$ for DNA or RNA and $A = 20$ for proteins), 
$\tau_{ik} = 1$ (0) if the allele at site $i$ is (is not)  
$k$, and the $\vec{\xi}_{ik}$ are uncorrelated random vectors. 
Clearly, our results for pairwise epistasis remain the same 
for this generalized model because they only concern 
mutations at different sites. However, the condition for 
a local fitness maximum now involves mutations to different alleles at the same site,
which may lead to a nontrivial dependence on $A$. 
On the basis of a recent study of evolutionary accessibility in
multiallelic sequence spaces, one may expect the fitness landscapes to
become less rugged with increasing $A$ \citep{Zagorski2016}, but this conjecture would have
to be corroborated by a detailed analysis.

Finally, whereas the present work focused on the structure of
the fitness landscapes induced by FGM, it is of obvious importance
to understand how the adaptive process actually proceeds on such a landscape
\citep{AO2005}. A simple framework that allows us to address this question
is provided by adaptive walks following Gillespie's strong
selection/weak mutation dynamics \citep{G1983,G1984,O2002}.
In a pioneering study, \cite{O1998} considered adaptive walks in FGM
assuming that the number $L$ of possible mutations is unlimited. 
In this setting, any population not located precisely at the
phenotypic optimum has a nonzero probability of generating another
beneficial mutation and the adaptive walk never stops; see
\cite{PK2008} for a related analysis of adaptation in the house-of-cards landscape.
For finite but large $L$, an interesting question concerns  
the number of steps until the population finds a local fitness maximum
when the adaptive dynamics is random \citep{KL1987,PSNK2015, PK2016} or greedy \citep{O2003,PNK2016}.
This problem is currently under investigation.

\section*{Acknowledgments}
We thank Anton Bovier, David Dean, Guillaume Martin, Olivier Tenaillon
and an anonymous reviewer for helpful remarks. 
This work was supported by Deutsche Forschungsgemeinschaft within Sonderforschungsbereich 680 ``Molecular Basis
of Evolutionary Innovations'' and Schwerpunktprogramm 1590 ``Probabilistic Structures in Evolution.'' S-CP acknowledges the support by the Basic Science Research Program through the National Research Foundation of Korea funded by the Ministry of Science, ICT and Future Planning (grant no. 2014R1A1A2058694). 
S-CP would also like to thank Korea Institute for Advanced Study
for its support and hospitality during his stay there on sabbatical leave (2016-2017).

\bibliography{me}

\appendix
\onecolumn
\newcommand{\rr}{\chi}
\newcommand{\Ln}{\mathrm{Ln}}
\renewcommand{\thesection}{Appendix~\Alph{section}}
\counterwithin*{equation}{section}
\setcounter{equation}{0}
\renewcommand{\theequation}{\Alph{section}\arabic{equation}}

\section{Derivation of the Joint Probability Density \boldmath{$ \mathcal{P}(R_1, R_2, R)$}}
\label{appendixA:DerivationOfJoinProb}
For the purpose of this calculation it will turn out to be convenient
to locate the wild-type phenotype on the diagonal of the trait space,
\textit{i.e.}, to set $\vec{Q} = \frac{Q}{\sqrt{n}}(1,1,1,\dots,)$. 
The probability density $\JointProb$ can then be formally defined as
\begin{align}
	\JointProb = n^3 
	\left \langle
	\delta\left\{ n R_1 - \sum_{i=1}^n \left [ (\xi_i + Q_i)^2 - Q_i^2\right ]\right\}   	
	\delta\left\{n R_2 - \sum_{i=1}^n \left [ (\eta_i + Q_i)^2 - Q_i^2\right ]\right\}   	
	\delta\left\{n R - \sum_{i=1}^n \left [ (\xi_i+\eta_i + Q_i)^2 - Q_i^2\right ]\right\}   	
	\right \rangle_{\vec{\xi}, \vec{\eta}},
	\label{CentralIdentity}
\end{align}
where $Q_i = Q/\sqrt{n}$ and $\Avr{\cdots}_{\vec{\xi},\vec{\eta}}$ stands for the average over the distribution of $ \vec{\xi} $ and $ \vec{\eta} $.
Using the integral representation of the $\delta$ function, 
we can write
\begin{align}
\JointProb	=\frac{n^3}{(2\pi)^3} 
	\int d\vec{k} e^{i  k_1 nR_1 + i  k_2 nR_2 + i  kn R}
		\prod_i \left \langle e^{- i k_1 (\xi_i + Q_i)^2 
			- i k_2 (\eta_i + Q_i)^2
- i k (\xi_i + \eta_i + Q_i)^2 +i Q_i^2 (k_1 + k_2 + k_3)}
	\right \rangle_{\vec{\xi}, \vec{\eta}},
\label{Eq:ID}
\end{align}
where $d\vec{R}$ and $d\vec{k}$ stand for $dR_1 dR_2 dR$ and $dk_1 dk_2 dk$, respectively, and we factorized the average by taking into account that 
the $\xi_i$'s and $\eta_i$'s are all independent and identically
distributed. 
The average in \eqref{Eq:ID} is readily calculated as
\begin{align}
	& \frac{1}{2\pi}\int _{-\infty }^{\infty }d\xi\int _{-\infty }^{\infty }  d \eta 
\exp \left [ 
-\frac{\eta^2}{2} 
-\frac{ \xi^2}{2}
- i k_1 \xi^2 - i k_2 \eta^2 - i k (\xi+\eta)^2 
- 2 i k_1 \frac{Q}{\sqrt{n}} \xi
- 2 i k_2 \frac{Q}{\sqrt{n}} \eta
- 2 i k \frac{Q}{\sqrt{n}} (\xi+\eta)
\right ]
\nonumber \\
	 =& \sqrt{\frac{1}{(1+ 2i k_1)(1+2 ik_2) - 4 k ( k_1+ k_2-i)}} 
\exp \left [ \frac{2 Q^2}{n} 
			 \frac{
2 i (k+k_1)(k+k_2)(k_1+k_2-i) + (k_1-k_2)^2
}
{4 k (k_1+k_2-i)+(2 k_1-i) (2 k_2-i)}
		 \right ],
\end{align}
which gives
\begin{align}
	\JointProb = \frac{n^3}{(2\pi)^3} 
		\int d\vec{k} \frac{e^{i  k_1 nR_1 + i  k_2 nR_2 + i  kn R}}{[(1+ 2i k_1)(1+2 ik_2) - 4 k ( k_1+ k_2-i) ]^{n/2}} 
\exp \left [ \frac{ n^2}{2 x^2} 
			 \frac{
2 i (k+k_1)(k+k_2)(k_1+k_2-i) + (k_1-k_2)^2
}
{4 k (k_1+k_2-i)+(2 k_1-i) (2 k_2-i)}
		 \right ]
			.
\end{align}	
In the limit $n \to \infty$, the integral is dominated by
contributions from the vicinity of the extremum of the exponent, which can be algebraically determined to be $ k= k_1= k_2=0$.
By expanding the argument of the exponential function up to the second
order around 
this point and performing the Gaussian integral, we obtain
\begin{align}
\JointProb
 \approx \frac{n^3}{(2\pi)^3} \int d \vec{k} 
\exp \Biggl \{ &- \frac{n^2}{2 x^2} \left [ (k_1 + k_2 + k)^2 - 2 k_1 k_2 
\right ] \nonumber \\
&- n \left [
4k^2 +k_1^2 + k_2^2 - i k_1 (R_1 -1) - i k_2 (R_2 -1) + k (2 i + 2 k_1 + 2 k_2 - i R)
\right ]
\Biggr \}\nonumber \\
=
\frac{\sqrt{n} x^2}{4\sqrt{2} \pi^{3/2}} [ 1 + O(&n^{-1})] \exp
\left \{ - \frac{n}{8} (R - R_1 - R_2)^2 - \frac{x^2}{2} \left [ (R_1-1)^2 + (R_2-1)^2 \right ] \right \},
\end{align}
which is \eqref{JointProb}.

\section{\label{appendixB:AsymptoticSignProb}Probability of Sign Epistasis}
In this appendix, we present the mathematical details of the
derivation of the probabilities $P_r$ and $P_s$ 
of observing RSE and SSE, respectively.
As in the main text, let us assume $R_1<R_2$.
In calculating the probabilities, 
the integral over $R$ takes one of three forms
\begin{align}
	\int_{-\infty }^{R_1} \sqrt{\frac{n}{8 \pi }} e^{ -\frac{n}{8} (-R+R_1+R_2)^2} \, dR  
		&= \frac{1}{2} \text{erfc}\left(\frac{\sqrt{n} R_2}{2 \sqrt{2}}\right),\qquad
	\int_{R_2}^{\infty} \sqrt{\frac{n}{8 \pi }} e^{ -\frac{n}{8} (-R+R_1+R_2)^2} \, dR 
		= \frac{1}{2} \text{erfc}\left(-\frac{\sqrt{n} R_1}{2 \sqrt{2}}\right) ,\text{ or}\nonumber \\
\int_{R_1}^{R_2} \sqrt{\frac{n}{8 \pi }} e^{ -\frac{n}{8} (-R+R_1+R_2)^2} \, dR &=\frac{1}{2} \left [ \text{erfc}\left(\frac{\sqrt{n} R_1}{2 \sqrt{2}}\right)
- \text{erfc}\left(\frac{\sqrt{n} R_2}{2 \sqrt{2}}\right) \right ].
\end{align}

First, we consider RSE which corresponds to the two domains
\begin{align}
D_1 = \{(R_1,R_2,R)| R_1<R_2<R, R_2<0\} \quad \text{ and }\quad
D_2 = \{(R_1,R_2,R)| R<R_1<R_2, R_1>0\},
\end{align}
as illustrated in Figure~\ref{fig:RegionForEachTypeOfEpistasis}. 
The probability of being in $D_1$ is
\begin{align}
\text{Pr}(D_1)&=\int_{-\infty}^0 dR_1 \int_{R_1}^0 dR_2 \int_{R_2}^\infty dR 
\mathcal{P}(R_1,R_2,R_3)
= \frac{x^2}{2\pi} \int_{-\infty}^0 dR_1 \int_{R_1}^0 dR_2
\exp\left [ -\frac{x^2}{2}(R_1-1)^2 -\frac{x^2}{2}(R_2-1)^2 \right ] 
\text{erfc}\left ( -\frac{\sqrt{n}R_1}{2\sqrt{2}} \right )\nonumber \\
&= \frac{x^2}{2\pi} \int_0^{\infty} dR_1 \int_0^{R_1} dR_2
\exp\left [ -\frac{x^2}{2}(R_1+1)^2 -\frac{x^2}{2}(R_2+1)^2 \right ] 
\text{erfc}\left ( \frac{\sqrt{n}R_1}{2\sqrt{2}} \right ),
\end{align}
where we have changed variables $R_i \mapsto -R_i$.
Since $\text{erfc}(y) \sim e^{-y^2}/(y \sqrt{\pi})$ for $y \gg 1$,
the above integral is dominated by the region $R_1  \ll 1$ for large $n$.
Thus, it is sufficient to approximate $\exp[-x^2((R_1-1)^2 + (R_2-1)^2)/2] 
\approx e^{-x^2}$, which yields 
\begin{align}
\text{Pr}(D_1) \approx \frac{x^2e^{-x^2}}{4\pi} \int_0^\infty dR_1 \int_0^{R_1} dR_2
\text{erfc}\left ( \frac{\sqrt{n}R_1}{2\sqrt{2}} \right )
\approx \frac{x^2e^{-x^2}}{4\pi} \int_0^\infty dR_1 
R_1 \text{erfc}\left ( \frac{\sqrt{n}R_1}{2\sqrt{2}} \right )
= \frac{2x^2e^{-x^2}}{n\pi} \int_0^\infty dy \, y \text{erfc}(y) = \frac{x^2}{2 n \pi} e^{-x^2}.
\end{align}
The probability of being in $D_2$ has the same leading behavior,
\begin{align}
\text{Pr}(D_2) &=\int_0^{\infty} dR_2 \int_0^{R_2} dR_1 \int_{-\infty}^{R_1} dR 
\mathcal{P}(R_1,R_2,R_3) 
= \frac{x^2}{2\pi} \int_0^{\infty} dR_2 \int_0^{R_2} dR_1
\exp\left [ -\frac{x^2}{2}(R_1-1)^2 -\frac{x^2}{2}(R_2-1)^2 \right ] 
\text{erfc}\left ( \frac{\sqrt{n}R_2}{2\sqrt{2}} \right )\nonumber \\
&\approx \frac{x^2}{2\pi} \int_0^{\infty} dR_1 \int_0^{R_1} dR_2
e^{-x^2}
\text{erfc}\left ( \frac{\sqrt{n}R_1}{2\sqrt{2}} \right ) =\frac{x^2}{2 n \pi} e^{-x^2},
\end{align}
where we have exchanged the variables $R_1 \leftrightarrow R_2$.
Due to the symmetrical roles of $R_1$ and $R_2$, the total probability of RSE is
\begin{align}
P_r \equiv 2 \sum_{i=1}^2 \text{Pr}(D_i) \approx \frac{2x^2}{n\pi} e^{-x^2}.
\end{align}

We can use a similar approximation scheme to calculate the probability of
SSE. There are four domains contributing to SSE (see Figure~\ref{fig:RegionForEachTypeOfEpistasis}), 
\begin{align}
D_3 &= \{(R_1,R_2,R)| R<R_1<0<R_2\},\quad
D_4 = \{(R_1,R_2,R)| R_1<0<R_2<R\},\nonumber \\
D_5 &= \{(R_1,R_2,R)| R_1<R<R_2<0\},\quad
D_6 = \{(R_1,R_2,R)| 0<R_1<R<R_2\}.
\end{align}
As we will see, all integrals can be represented by the functions
\begin{align}
G_1(a,b) &= \frac{x^2}{4 \pi} \int_0^\infty dR_1 \int_0^{R_1} dR_2
\exp\left [ -\frac{x^2}{2}(R_1+a)^2 -\frac{x^2}{2}(R_2+b)^2 \right ] 
 \text{erfc}\left ( \frac{\sqrt{n}R_2}{2\sqrt{2}}\right ) ,\\
G_2(a,b) &= \frac{x^2}{4 \pi} \int_0^\infty dR_1 \int_0^{R_1} dR_2
\exp\left [ -\frac{x^2}{2}(R_1+a)^2 -\frac{x^2}{2}(R_2+b)^2 \right ] 
 \text{erfc}\left ( \frac{\sqrt{n}R_1}{2\sqrt{2}}\right ) \nonumber \\
&=  \frac{x}{4 \sqrt{2 \pi}} 
 \text{erfc}\left ( \frac{bx}{\sqrt{2}}\right )
\int_0^\infty dR_1 
\exp\left [ -\frac{x^2}{2}(R_1+a)^2 \right ] 
 \text{erfc}\left ( \frac{\sqrt{n}R_1}{2\sqrt{2}}\right )
 - G_1(b,a),
\end{align}
where $a, b = \pm 1$ and we have used that
\begin{align}
\int_0^\infty dy \int_{y}^\infty dz f(y,z) 
= \int_0^\infty dy \int_0^{y} dz f(z,y).
\label{Eq:int_ch}
\end{align}
To be specific, we write the probabilities of being in each domain as
\begin{align*}
\text{Pr}(D_3) &= 
 \frac{x^2}{4\pi} \int_{-\infty}^0 dR_1 \int_0^\infty dR_2
\exp\left [ -\frac{x^2}{2}(R_1-1)^2 -\frac{x^2}{2}(R_2-1)^2 \right ] 
\text{erfc}\left ( \frac{\sqrt{n}R_2}{2\sqrt{2}}\right )
= G_1(1,-1) + G_2(-1,1),\\
\text{Pr}(D_4) &= \frac{x^2}{4\pi} \int_{-\infty}^0 dR_1 
\int_0^\infty dR_2
\exp\left [ -\frac{x^2}{2}(R_1-1)^2 
 -\frac{x^2}{2}(R_2-1)^2 \right ] 
\text{erfc}\left ( -\frac{\sqrt{n}R_1}{2\sqrt{2}}\right )
= G_1(-1,1) + G_2(1,-1),\\
\text{Pr}(D_5) &= \frac{x^2}{4\pi} \int_{-\infty}^0 dR_1 \int_{R_1}^0 dR_2
\exp\left [ -\frac{x^2}{2}(R_1-1)^2 -\frac{x^2}{2}(R_2-1)^2 \right ] 
\left [ \text{erfc}\left ( \frac{\sqrt{n}R_1}{2\sqrt{2}}\right )- \text{erfc}\left ( \frac{\sqrt{n}R_2}{2\sqrt{2}} \right )\right ] = G_1(1,1)-G_2(1,1),
\\
\text{Pr}(D_6) &= \frac{x^2}{4\pi} \int_0^{\infty} dR_1 \int_{R_1}^\infty dR_2
\exp\left [ -\frac{x^2}{2}(R_1-1)^2 -\frac{x^2}{2}(R_2-1)^2 \right ] 
\left [ \text{erfc}\left ( \frac{\sqrt{n}R_1}{2\sqrt{2}}\right )- \text{erfc}\left ( \frac{\sqrt{n}R_2}{2\sqrt{2}} \right )\right ] = G_1(-1,-1)-G_2(-1,-1),
\end{align*}
where we have changed negative integral domains into positive domains and
made use of \eqref{Eq:int_ch}.
Using 
the approximation scheme explained above, we get 
\begin{align}
G_1(a,b) 
&= \frac{x}{4\sqrt{2 \pi}} \int_0^\infty dR_2 
e^{-x^2(R_2+b)^2/2}
\text{erfc}\left ( \frac{\sqrt{n}R_2}{2\sqrt{2}}\right ) 
\text{erfc}\left [ \frac{x(R_2+a)}{\sqrt{2}}\right ] 
\nonumber \\
&\approx \frac{x}{4\sqrt{2 \pi}} \text{erfc} \left ( \frac{ax}{\sqrt{2}} \right ) e^{-x^2/2}
\int_0^\infty dR_2 \text{erfc}\left ( \frac{\sqrt{n}R_2}{2\sqrt{2}}\right ) 
= \frac{x}{2 \sqrt{n} \pi} \text{erfc} \left ( \frac{ax}{\sqrt{2}} \right ) e^{-x^2/2} + O(1/n).
\end{align}
Since
\begin{align}
\frac{x}{4 \sqrt{2 \pi}} 
 \text{erfc}\left ( \frac{bx}{\sqrt{2}}\right )
\int_0^\infty dR_1 
\exp\left [ -\frac{x^2}{2}(R_1+a)^2 \right ] 
 \text{erfc}\left ( \frac{\sqrt{n}R_1}{2\sqrt{2}}\right )
\approx \frac{x}{2 \sqrt{n} \pi} \text{erfc} \left ( \frac{bx}{\sqrt{2}} \right ) e^{-x^2/2} + O(1/n),
\end{align}
we conclude that $G_2(a,b) = O(1/n)$.
Using $\text{erfc}(y) + \text{erfc}(-y) = 2$, we finally obtain
\begin{align}
P_s \equiv 2 \sum_{i=3}^6 \text{Pr}(D_i) \approx \frac{4x}{\sqrt{n} \pi} e^{-x^2/2}.
\end{align}

\section{\label{sec:DerivProbLocalOptima}Large \boldmath{$L$} Behavior of \boldmath{$\mathcal{R}_s(L)$} for Fixed \boldmath{$n$}}
In this appendix, we calculate the asymptotic behavior of the
probability $\mathcal{R}_s(L)$ for a genotype with $s$ mutations to be a local
fitness maximum in the limit where $L$ is large and the phenotype
dimensions $n$ is fixed. As explained in the main text, this
probability has two contributions which arise from expanding the
function $F(\vec{k},\vec{z})$ defined in \eqref{Fdef} near $\vert \vec{z} \vert = 0$ 
and $\vert \vec{z} \vert = z^\ast \sim L$, respectively (see
\eqref{Rsdecomp}). 

First, we consider the contribution from the region $|\vec{z} | \ll 1$.
In this case, we can approximate $F(\vec{k},\vec{z})$ as
\begin{align}
F(\vec{k},\vec{z}) = 
\int_n e^{-i \vec{k}\cdot \vec{\xi}} p(\vec{\xi}) d\vec{\xi} - 
\int_{\mathcal{D}^c(\vec{z})} e^{-i \vec{k}\cdot \vec{\xi}} p(\vec{\xi}) d\vec{\xi} 
\approx e^{-k^2/2}   - \int_{\mathcal{D}^c(\vec{z})} p(0) d\vec{\xi}
= e^{-k^2/2} - A_n z^n 
\approx \exp \left [ - \frac{k^2}{2} - A_n z^n e^{k^2/2} \right ],
\end{align}
where $k = |\vec{k}|$; $z = |\vec{z}|$; $\mathcal{D}^c(\vec{z}) = \{\vec{y} | |\vec{y} - \vec{z}| \le z \}$,
which is the complement of $\mathcal{D}(\vec{z})$; $A_n = p(0) S_{n-1}/n$ with $S_{n-1} = 2 \pi^{n/2}/\Gamma(n/2)$ being the surface area
of the unit sphere in $(n-1)$ dimensions; and $p(0) = (2\pi)^{-n/2}$. Note that the error of the
above approximation is $O(z^{n+1})$.
Thus, setting $\rho \equiv s/L$ we can approximate
\begin{align}
\mathcal{R}_s^<(L) \approx \int_n \frac{d\vec{z} d\vec{k}}{(2\pi)^n} 
\exp\left [ i \vec{k} \cdot \vec{z} + L H_1(\vec{k},\vec{z})
\right ], \quad
H_1(\vec{k},\vec{z}) \equiv -i \vec{k} \cdot   \vec{q}  - \rho \frac{k^2}{2}
- \rho A_n z^n e^{k^2/2} -(1-\rho)  A_n z^n ,
\end{align}
where $\vec{q} = \vec{Q}/L$.
Since $L$ is large, we can employ the saddle point approximation. 
One can easily see that the saddle point solving the equations $\partial_{k_j} H_1 = \partial_{z_k} H_1 = 0$
is at $\vec{z} = 0$ and $k_j = -i q_j/\rho$. Around the saddle point, we expand
\begin{align}
H_1 \approx -\frac{q^2}{2 \rho} - \frac{\rho}{2} \left ( \vec{k} + i \vec{q}/\rho\right )^2 - A_n z^n \left [ \rho e^{-q^2/(2 \rho^2)} + (1-\rho) \right ],
\end{align}
which gives
\begin{align}
\label{Rsapp}
\mathcal{R}_s^<(L) &\approx \exp\left ( -L \frac{q^2}{2\rho} \right )
\int_n d\vec{z} \exp \left [ -L A_n z^n \left \{ \rho e^{-q^2/(2 \rho^2)} + (1-\rho) \right \} \right ] \int_n \frac{d\vec{k}}{(2\pi)^n} 
\exp \left [ - \frac{L \rho}{2} \left ( \vec{k} + i \vec{q}/\rho\right )^2
\right ] \nonumber \\
&= \frac{\exp\left ( -L \frac{q^2}{2\rho} \right )}{(2 \pi L \rho)^{n/2}}
\int_0^\infty S_{n-1} z^{n-1} dz \exp \left [ -L A_n z^n\left \{ \rho e^{-q^2/(2 \rho^2)} + (1-\rho) \right \} \right ]
= 
\frac{s^{-n/2}\exp\left [ - Q^2/(2s) \right ] }{s \exp[-Q^2/(2s^2)] + L-s}
\left [ 1 + O(L^{-1/n})\right ],
\end{align}
and the last step involves a change of variables $z \to t=S_{n-1} z^n /n$.
Since $L$ appears in the integrand in the combination $L z^n$, the
error that arises from neglecting terms of $O(z^{n+1})$ is
$L^{-1/n}$. The leading order of \eqref{Rsapp} was reported in \eqref{RSFirstPhase}. 

Now we move on to the calculation of $\mathcal{R}_s^>(L)$, where the
dominant contribution to $F(\vec{k},\vec{z})$ comes from a region where $z \sim O(L)$.
Using $\int d\vec{\xi} p(\vec{\xi}) \exp(-i \vec{k} \cdot \vec{\xi} )
= \exp(-k^2/2)$, we calculate the integral $I \equiv \exp(-k^2/2)- F(\vec{k},\vec{z})$ 
as
\begin{align}
I
&= \frac{1}{(2\pi)^{n/2}} \int_0^{2z} d\xi_n e^{- i k_n \xi_n-\xi_n^2/2} 
\int_{B(2z,\xi_n)} d \vec{\xi}_\perp e^{-i \vec{k}_\perp \cdot
\vec{\xi}_\perp-\vec{\xi}_\perp^2/2}\nonumber \\
&= \frac{1}{(2\pi)^{n/2}} \int_0^{2z} d\xi_n e^{- i k_n \xi_n-\xi_n^2/2} 
\left [ \int_{\mathbb{R}^{n-1}} d \vec{\xi}_\perp e^{-i \vec{k}_\perp \cdot
\vec{\xi}_\perp-\vec{\xi}_\perp^2/2} - 
\int_{B^c(2z,\xi_n)} d \vec{\xi}_\perp e^{-i \vec{k}_\perp \cdot
\vec{\xi}_\perp-\vec{\xi}_\perp^2/2} \right ] \nonumber \\
&= \frac{e^{-\vec{k}_\perp^2/2}}{\sqrt{2\pi}} 
\left [ \int_0^\infty d \xi_n e^{-i  k_n \xi_n -  \xi_n^2/2} - \int_{2z}^\infty  d \xi_n e^{-i  k_n \xi_n -  \xi_n^2/2} \right ]
- \frac{1}{(2\pi)^{n/2}}\int_0^{2z} d\xi_n e^{- i k_n \xi_n-\xi_n^2/2}\int_{B^c(2z,\xi_n)} d \vec{\xi}_\perp e^{-i \vec{k}_\perp \cdot
\vec{\xi}_\perp-\vec{\xi}_\perp^2/2} \nonumber \\
&=\frac{e^{-k^2/2}}{2} \left ( 1 - \text{erf}\left ( \frac{i k_n}{\sqrt{2}} \right )\right ) -  \frac{e^{-\vec{k}_\perp^2/2}}{\sqrt{2\pi}} \int_{2z}^\infty  d \xi_n e^{-i  k_n \xi_n -  \xi_n^2/2} - C_1(\vec{k},z),
\end{align}
where we set $\vec{z} = z \vec{e}_n$,  
$\vec{\xi}_\perp
= \vec{\xi} - \xi_n \vec{e}_n$ and $\vec{k}_\perp = \vec{k} - k_n
\vec{e}_n$ with $\vec{e}_n = (0,\ldots,0,1)$,
$B(2z,\xi_n)$ is an $(n-1)$-dimensional ball with radius $\sqrt{\xi_n (2z - \xi_n)}$ whose center is located at the origin, $B^c$ is the relative complement of $B$ with respect to $\mathbb{R}^{n-1}$, and $\text{erf}(z) = 2 \int_0^z e^{-t^2} dt/\sqrt{\pi}$ is the error function. The definition of $C_1$ is self-explanatory.
Since 
\begin{align}
\left | \int_{2z}^\infty  d \xi_n e^{-i  k_n \xi_n -  \xi_n^2/2} \right |
\le \int_{2z}^\infty d \xi_n e^{-  \xi_n^2/2} 
= 2z \int_1^\infty d y e^{-2 z^2 y^2}
\approx \frac{e^{-2z^2}}{2 z},
\end{align}
where we used the Laplace method for the asymptotic expansion, 
the leading finite $z$ correction is expected to come from $C_1$ for $n>1$.
Note that  $C_1 $ is identically zero for $n=1$.
Thus we get
\begin{align}
\label{FC1}
F(\vec{k},\vec{z} ) \approx \frac{1}{2} e^{-k^2/2} \left ( 1 + \text{erf}\left ( \frac{i \vec{k}\cdot \vec{z}}{\sqrt{2}z} \right )
\right ) + C_1(\vec{k},z),
\end{align}
where $k_n$ is written as a projection of $\vec{k}$ along
the $\vec{z}$ direction, $k_n = \vec{k} \cdot \vec{z}/z$.
Since
\begin{align}
\left | C_1(\vec{k},z) \right | 
\le \frac{1}{(2\pi)^{n/2}}\int_0^{2z} d\xi_n e^{-\xi_n^2/2}\int_{B^c(2z,\xi_n)} d \vec{\xi}_\perp e^{-\vec{\xi}_\perp^2/2} =C_1(0,z) ,
\end{align}
it is sufficient to find an approximate formula for $C_1(0,z)$ to determine
the $z$ dependence of $C_1(\vec{k},z)$. 
Using spherical coordinates in $\mathbb{R}^{n-1}$, we get
\begin{align}
C_1(0,&z)= \frac{S_{n-2}}{(2\pi)^{n/2}}\int_0^{2z} dy e^{-y^2/2}
\int_{\sqrt{y(2z-y)}}^\infty d x x^{n-2} e^{-x^2/2}\nonumber \\
&=\frac{S_{n-2}}{(2\pi)^{n/2}} 
\left [ \int_0^{2z} dy e^{-y^2/2}\int_z^\infty dx x^{n-2} e^{-x^2/2}
+ \int_0^z dx x^{n-2} e^{-x^2/2} \left \{ \int_0^{M_-(x,z)} dy e^{-y^2/2}
+\int_{M_+(x,z)}^z dy e^{-y^2/2} \right \} \right ],
\end{align}
where $S_{n-2} = 2 \pi^{(n-1)/2}/\Gamma[(n-1)/2]$ is the surface 
area of the unit $(n-2)$ sphere. In the second term on the second line, the order
of integration was reversed and the integration boundaries $M_\pm(x,z)
= z \pm \sqrt{z^2-x^2}$ were introduced.
Since the first integral ($\int_z^\infty dz$) and the third integral 
($\int_{M_+}^z dy$) decrease exponentially with $z$,
the main contribution to $C_1(0,z)$ comes from the second integral. Thus,
\begin{align}
C_1(0,z) &\approx \frac{S_{n-2}}{(2\pi)^{n/2}} 
\int_0^z dx x^{n-2} e^{-x^2/2} \int_0^{M_-(x,z)} dy e^{-y^2/2}
= \frac{S_{n-2} z^n}{(2\pi)^{n/2}} 
\int_0^1 dx x^{n-2} e^{-z^2 x^2/2} \int_0^{M_-(x,1)}  dy e^{-z^2 y^2/2} \nonumber \\
&= \frac{S_{n-2} z^{n-1}}{(2 \pi)^{n/2} } \sqrt{\frac{\pi}{2}}
\int_0^1 dx x^{n-2} e^{-z^2 x^2/2} \text{erf} (M_-(x,1) z/\sqrt{2}).
\end{align}
Since the last integral is dominated by the region $xz \le 1$,
we can approximate $M_-(x,1)z \approx x^2 z/2 \sim O(1/z)$ and
$\text{erf} (M_-(x,1) z/\sqrt{2}) = x^2 z/\sqrt{2\pi}$.
Finally, we get
\begin{align}
\label{C10}
C_1(0,z) \approx \frac{S_{n-2} z^n}{(2 \pi)^{{n+1}/2} } \sqrt{\frac{\pi}{2}}
\int_0^1 dx x^n e^{-z^2 x^2/2} 
\approx \frac{S_{n-2} z^n}{(2 \pi)^{{n+1}/2} } \sqrt{\frac{\pi}{2}}
\int_0^\infty dx x^n e^{-z^2 x^2/2}
= \frac{n-1}{2 \sqrt{2\pi} z},
\end{align}
which also implies that $C_1(\vec{k},z) \sim O(z^{-1})$.

If we write
\begin{align}
\label{FC12}
F(\vec{k},\vec{z}) = \frac{1}{2} e^{-k^2/2} \left [ 1 +
  \text{erf}\left ( \frac{i \vec{k} \cdot \vec{r}}{\sqrt{2} r} \right )
\right ] \left [ 1 + \frac{1}{L} f(\vec{r},\vec{k}) +O(z^{-2})\right ],
\end{align}
where $\vec{r}=\vec{z}/L$ and $r=z/L$, then comparison with \eqref{FC1} and \eqref{C10}
shows that $f(\vec{r},0) = (n-1)/(\sqrt{2 \pi} r)$.  
Inserting \eqref{FC12} into \eqref{Rsdefs}, it follows that  
\begin{align}
\mathcal{R}_s^>(L) \approx \frac{L^n}{2^L} \int \frac{d\vec{r} d\vec{k}}{(2\pi)^n}
\exp \left [ \rho f(\vec{r},\vec{k})+ (1-\rho) f(\vec{r},0) \right ]
\exp \left [ L H_2(\vec{k},\vec{r}) \right ],
\end{align}
with 
\begin{align}
H_2(\vec{k},\vec{r}) =  i \vec{k}\cdot(\vec{r} - \vec{q} )
- \rho \frac{k^2}{2} +\rho \ln \left ( 1 + \text{erf}\left ( \frac{i \vec{k}\cdot\vec{r}}{\sqrt{2}r} \right ) \right ) .
\end{align}
Now we employ the steepest-descent method. For convenience, we set
$\vec{q} = (q,0,\ldots,0)$. The saddle point satisfies the equations
\begin{align}
\frac{\partial H_2}{\partial r_j} &= i k_j + i \frac{\sqrt{2}\rho}{\sqrt{\pi}r^3} \exp \left [ \frac{(\vec{k}\cdot \vec{r})^2}{2 r^2}  \right ]
\left [ k_j r^2 - r_j (\vec{k} \cdot \vec{r} ) \right ]\left [ 1 + \text{erf}\left ( \frac{i \vec{k}\cdot\vec{r}}{\sqrt{2}r} \right ) \right ]^{-1} = 0,\\
\frac{\partial H_2}{\partial k_j} &= i (r_j-q\delta_{j1}) - \rho k_j + i \frac{\sqrt{2}\rho}{\sqrt{\pi}r} \exp \left ( \frac{(\vec{k}\cdot \vec{r})^2}{2 r^2}  \right )
r_j \left ( 1 + \text{erf}\left ( \frac{i \vec{k}\cdot\vec{r}}{\sqrt{2}r} \right ) \right )^{-1} = 0,
\end{align}
with the solution $\vec{k}^* = 0$ and $\vec{r}^* = (q - \rho\sqrt{2/\pi},0,\ldots,0)$. Note that there is no solution if $q < \rho \sqrt{2/\pi} $,
so the valid range of $\rho$ has the upper boundary 
$\rho_c(q) = \min(1,\sqrt{\pi/2}q)$. The matrix of 
second derivatives around the saddle point $\vec{k}^*, \vec{r}^*$ is
\begin{align}
\left . \frac{\partial^2 H_2}{\partial r_l \partial r_j} \right |_* = 0,\quad
\left . \frac{\partial^2 H_2}{\partial k_m \partial k_j} \right |_* = 
- \rho \delta_{mj} \left ( 1 - \frac{2 }{\pi} \delta_{l1}\right ),\quad
\left . \frac{\partial^2 H_2}{\partial r_m \partial k_j} \right |_* = 
i \delta_{jl} \left [ 1 +  (1-\delta_{m1}) \frac{\sqrt{2}\rho} 
{\sqrt{\pi} q- \sqrt{2} \rho} \right ].
\end{align}
Thus, we get
\begin{align}
\mathcal{R}_s^>(L) \approx \frac{L^n}{2^L} e^{f(\vec{r}^*,0)}
\int \frac{d\vec{y} d\vec{k}}{(2\pi)^n}
\exp \left [ -L \frac{\rho}{\pi} (\pi - 2) k_1^2  - L \rho \vec{k}_\perp^2
+ i L k_1 y_1 + i L\frac{\sqrt{\pi} q}{\sqrt{\pi} q - \sqrt{2} \rho} \vec{k}_\perp \cdot \vec{y}_\perp \right ],
\end{align}
where $\vec{y}  = \vec{r} - \vec{r}^*$, 
$\vec{y}_\perp = (0,y_2,\ldots,y_n)$, and
$\vec{k}_\perp = (0,k_2,\ldots,k_n)$.
If we perform the integration over $\vec{y}$ first, we obtain $\delta$ functions
which make the integral over $\vec{k}$ trivial. Finally, we arrive at
\begin{align}
\mathcal{R}_s^>(L) \approx 2^{-L} \theta\left ( \rho_c - \rho \right )
\left ( \frac{ q-2 \rho/\sqrt{2\pi}}{ q}  
\exp \left ( \frac{1}{\sqrt{2\pi} q - 2 \rho} \right )
\right )^{n-1},
\end{align}
where $\theta(x)$ is the Heaviside step function defined by $\theta(x
\geq 0) = 1$ and $\theta(x < 0) = 0$.

To evaluate the corresponding contribution to the number of fitness
maxima, $\mathcal{N}_>$, we replace the summation over $s$ in
\eqref{Eq:Ngtlt} by an integral over $\rho = s/L$  
and use Stirling's formula to
approximate the binomial coefficients. This yields
\begin{align}
\mathcal{N}_> &= \sqrt{L} \int_0^1 d\rho \frac{\exp[ L \{- \rho\ln \rho - (1-\rho) \ln (1-\rho) - \ln 2\}]}{\sqrt{2 \pi \rho ( 1-\rho)}} 
\left ( \frac{ q-2 \rho/\sqrt{2\pi}}{ q}
\exp \left ( \frac{1}{\sqrt{2\pi} q - 2 \rho} \right )
\right )^{n-1} 
\theta(\rho_c - \rho). 
\end{align}
If $\rho_c < \frac{1}{2}$ or $q < \sqrt{2 \pi}^{-1} = q_0$, the integral is dominated around $\rho \approx
\rho_c$, which results in an exponential decrease with $L$. On the
other hand, if $\rho_c > 1/2$, the integral is dominated around $\rho \approx
1/2$, which gives
\begin{align}
\mathcal{N}_> \approx \sqrt{\frac{2L}{\pi}} \int_{-\infty}^\infty dx e^{-2Lx^2} 
\left [ \frac{ q-1/\sqrt{2\pi}}{ q} 
\exp \left ( \frac{1}{\sqrt{2\pi} q - 1} \right )
\right ]^{n-1}
= 
\left [ \frac{ q-1/\sqrt{2\pi}}{ q}
\exp \left ( \frac{1}{\sqrt{2\pi} q - 1} \right )
\right ]^{n-1}
\end{align}
as reported in \eqref{NumLocalMaximaNFixedSecond}. 

\section{Derivation of \eqref{ReducedComplexity}}
\label{appendixF:JoinLimit}
In this appendix, we calculate the average number of fitness maxima
$\NumMax$ in the limit $n, L \rightarrow \infty$ at fixed ratio $\alpha \equiv n/L$.
To this end, we write $I_\tau$, the probability for the genotype
$\tau$ to be a local fitness maximum, using the Heaviside step function as
\begin{align}
I_\tau 
= \int \prod_{k=1}^L \left [ d\vec{\xi}_k p(\vec{\xi}_k) 
\theta\left( \frac{1}{L}(\vec{z} +(1-2\tau_k) \vec{\xi}_k )^2 - \frac{1}{L} |\vec{z}|^2 \right) \right ]
\equiv \int \mathcal{D}\xi \prod_k 
\theta\left ( \mathcal{E}_k \right ),
\label{Eq:Itau_int}
\end{align}
where 
$\vec{z}$ is determined by $\tau$ through \eqref{Eq:ztau}, 
$\int \Measure{\xi} \equiv \int \prod_k d \vec{\xi}_k p(\vec{\xi}_k)$, 
and $\mathcal{E}_k$ is defined as
\begin{align}
\mathcal{E}_k &= \frac{1}{L} \left(
	\vec{z} + \vec{\xi}_k (1 - 2 \tau_k)
\right)^2 - \frac{1}{L} |
	\vec{z}
|^2 = \frac{1}{L} \left ( 2 \vec{z} \cdot \vec{\xi}_k (1 - 2 \tau_k) + 
\left | \vec{\xi} \right |^2 \right ) 
= \frac{1}{L} \left[
	2 \left(\vec{Q} + \sum_{j} \vec{\xi}_j \tau_j \right) \cdot \vec{\xi}_k (1 - 2 \tau_k) + |\vec{\xi}_k|^2 
\right]. 
\label{EnergyDifference}
\end{align}
Note that the prefactor $ 1/L $ is introduced to make $\mathcal{E}_k$  finite 
in the limit $L\to\infty$ and we have used that $(1-2\tau_k)^2 = 1$.
Applying the identity \citep{Tanaka1980, Bray1980}
\begin{align}
\theta(\mathcal{E}_k) = \int_0^\infty d\lambda_k \delta
\left (\lambda_k - \mathcal{E}_k \right )
= \int_0^\infty d\lambda_k \int_{-\infty}^\infty \frac{d\phi_k}{2\pi}
\exp\left [ i \phi_k \left ( \lambda_k - \mathcal{E}_k \right ) \right ]
\end{align}
to \eqref{Eq:Itau_int}, the expected number of local fitness maxima reads
\begin{align}	
	\NumMax &= \Tr \int \Measure{\xi} \prod_{k=1}^L 
\left [ \int_{0}^{\infty} d\lambda_k \int_{-\infty}^{\infty} \frac{d \phi_k}{2 \pi} 
		e^{i \phi_k(\lambda_k - \mathcal{E}_k)} \right ] 
= \Tr \int \Measure{\xi} \Measure{\lambda} \Measure{\phi} 
\exp\left [ 
		\sum_{k=1}^L \left \{ i \phi_k \lambda_k + \frac{i}{L} \phi_k \left(
		2 \vec{\xi}_k  \cdot \sum_{j=1}^L \vec{\xi}_j \tau_j  + 2 \vec{\xi}_k \cdot \vec{Q} - |\vec{\xi}_k|^2
		\right) \right \}
\right ]
\nonumber \\
	&= \Tr \int \Measure{\xi} \Measure{\lambda} \Measure{\phi} 
\exp\left [ 
		\sum_{k=1}^L \left \{ i \phi_k \lambda_k + \frac{i}{L} \phi_k \left(
		 2 \vec{\xi}_k \cdot \vec{Q} - |\vec{\xi}_k|^2
		\right) \right \}
\right ]\exp \left [ \frac{i}{L}\sum_{k=1}^L  \phi_k \vec{\xi}_k  \cdot \sum_{j=1}^L \vec{\xi}_j \left ( 2 \tau_j \right )
\right ],
	\label{NumMetaStableFormal}	
\end{align}
where $\int \Measure{\lambda} \equiv \int_{0}^{\infty} \prod_k d \lambda_k $,
	$ \int \Measure{\phi} \equiv \int_{-\infty}^{\infty} \prod_k
        d \phi_k/2 \pi $, and we made the change of variables $(2\tau_k-1) \vec{\xi}_k \mapsto \vec{\xi}_k  $ to arrive at the second equality. 
Using the identity 
\begin{align}
\exp\left( \frac{i}{L} \vec{X} \cdot \vec{Y} \right) = 
L^n\int_n d\vec{\nu} \delta\left ( L \vec{\nu} - \vec{X} \right )
\exp\left ( i \vec{Y} \cdot \vec{\nu} \right )
= 
\left ( \frac{L}{2\pi} \right )^n \int_n d\vec{\mu} d\vec{\nu} \exp \left[
i L \vec{\mu}\cdot \vec{\nu} - i \vec{X}\cdot \vec{\mu} + i \vec{Y}\cdot \vec{\nu}
\right],
\label{HubbardStratonovich}
\end{align}
which is valid for any $n$-dimensional real vectors $\vec{X}$ and $\vec{Y}$,
we can write the last term of \eqref{NumMetaStableFormal} as
\begin{align}
\exp \left[
	\frac{i}{L} \sum_{k} \phi_k \vec{\xi}_k \cdot \sum_j \vec{\xi}_j (2\tau_j)
\right] 
= \left(\frac{L}{2\pi}\right)^{n} \int_n d\vec{\mu} d\vec{\nu} 
\exp \left [
	i L \vec{\mu} \cdot \vec{\nu} + i \sum_{k} \vec{\xi}_k \cdot (- \phi_k \vec{\mu} + 2 \tau_k \vec{\nu})
\right ]
,
\end{align}
which gives
\begin{align}
\NumMax 
= \Tr \int  \Measure{\lambda} \Measure{\phi} \Measure{\mu}
\Measure{\nu} e^{i \phi \cdot \lambda + i L \vec{\mu} \cdot \vec{\nu}}
\prod_{k=1}^L\prod_{s=1}^n \int \frac{d\xi_{ks}}{\sqrt{2 \pi}}
\exp\left [ 
		i \left \{  \frac{1}{L} \phi_k \left(
		 2 \xi_{ks} Q_s - \xi_{ks}^2
		\right)+ {\xi}_{ks} (- \phi_k \mu_s + 2 \tau_k \nu_s)\right \} - \frac{\xi_{ks}^2}{2}
\right ],
\label{Eq:Num}
\end{align}
where $\int \Measure{\mu}\Measure{\nu} \equiv  \left(\frac{L}{2\pi}\right)^{n} 
\int_n d\vec{\mu} d\vec{\nu}$, $\phi \cdot \lambda \equiv 
\sum_{k=1}^L \phi_k \lambda_k$, and
$\xi_{ks}$, $Q_s$, $\mu_s$, $\nu_s$ are the $s$-th components of the vectors
$\vec{\xi}_k$,
$\vec{Q}$,
$\vec{\mu}$,
$\vec{\nu}$, respectively.
Note that the integrals over the $\xi_{ks}$'s become independent of each other.
If we choose 
$Q_s = Q/\sqrt{n} = q \sqrt{L/\alpha} $ for all $s$ and define $\rr \equiv 2Q_s/L$, 
the integral over $\xi_{ks}$ becomes
\begin{align}
	\int_{-\infty}^\infty & \frac{d\xi_{ks}}{\sqrt{2 \pi }} 
\exp\left [ - \frac{\xi_{ks}^2}{2} \left ( 1 + 2 i \phi_k /L \right ) + 
i {\xi}_{ks} \left \{ \phi_k (\rr - \mu_s) + 2 \tau_k \nu_s \right \}
\right ] 
	 = \frac{1}{\sqrt{1 + 2 i \phi_k /L}} 
\exp\left [
		-\frac{\left \{2\nu_s  \tau_k +(\rr- \mu_s)  \phi_k \right \}^2 }{2(1 + 2 i \phi_k /L)} 
	\right ] \nonumber \\
	&= \frac{1}{\sqrt{1 + 2 i \phi_k /L}} 
\exp\left [
		\frac{-(\mu_s-\rr)^2  \phi_k^2 +4 \tau_k (\mu_s-\rr) \nu_s \phi_k -4\nu_s^2  \tau_k }{2(1 + 2 i \phi_k /L)} 
	\right ],
\end{align}
which, in turn, gives
\begin{align}
\NumMax = 
 \Tr \int  \Measure{\lambda} \Measure{\phi} \Measure{\mu}
\Measure{\nu} e^{i \phi \cdot \lambda + i L \vec{\mu} \cdot \vec{\nu}}
\prod_{k=1}^L\frac{1}{(1 + 2 i \phi_k /L)^{n/2}} 
\exp\left [
		\frac{-\phi_k^2\sum_s(\mu_s-\rr)^2   +4 \tau_k\phi_k \sum_s (\mu_s-\rr) \nu_s  -4\tau_k \sum_s \nu_s^2  }{2(1 + 2 i \phi_k /L)} 
	\right ].
	\label{NumMaxBeforeLagrange}
\end{align}
If we now insert the identity 
\begin{align}
1 &= \int_0^\infty da \int db dc \delta \left ( a - \sum_s(\mu_s-\rr)^2 \right )
\delta \left ( b - \sum_s(\mu_s-\rr)\nu_s \right )
\delta \left ( c - \sum_s\nu_s^2 \right )\nonumber \\
&=\int_{0}^{\infty} da \int db dc \int \frac{dA}{2\pi / L } \frac{dB}{2\pi / L } \frac{dC}{2\pi / L }  
\exp \left [ 
	i A L \left\{
		 a - \sum_s (\mu_s - \rr)^2
	\right\}
	+ i B L \left\{
		b - \sum_s (\mu_s - \rr) \nu_s
	\right\}
	+ i C L \left\{
		c - \sum_s \nu_s ^2
	\right\}
\right ],
\label{LagrangeMultipliers}
\end{align}
we can write
\begin{align}
	\NumMax =& \Tr \int \Measure{\lambda} \Measure{\phi} \Measure{\mu}\Measure{\nu} e^{i \phi \cdot \lambda + i L \vec{\mu} \cdot \vec{\nu}}\int_0^\infty da \int \frac{dA db dB dc dC}{(2\pi / L)^3 }  
		 \prod_k (1 + 2 i \phi_k /L)^{-L\alpha/2}  
\exp \left [
			\frac{ -a \phi_k^2 + 4 b \tau_k \phi_k - 4 c \tau_k}{2(1 + 2 i \phi_k /L)}  
	\right ] 
\nonumber \\ 
&\times \exp \left \{ 
	i A L \left[
		 a - \sum_s (\mu_s - \rr)^2
	\right]
	+ i B L \left[
		b - \sum_s (\mu_s - \rr) \nu_s
	\right]
	+ i C L \left[
		c - \sum_s \nu_s ^2
	\right]
\right \}
\end{align}
where we have replaced $n$ by $L \alpha$. The integral domain of $a$
is restricted to the positive real axis to ensure that the integral with respect to $\phi_k$ in \eqref{NumMaxBeforeLagrange} continues to be well-defined after the substitution. 
Performing the integrals over $\mu_s$ and $\nu_s$, we get
\begin{align}
	\frac{L}{2\pi}  \int d \mu_s d \nu_s
	e^{
		i L \mu_s \nu_s - iL A (\mu_s - r)^2  - i L B(\mu_s -r ) \nu_s - i L C  \nu_s^2
	} 
	&=\exp\left[
	 -\frac{1}{2} \left(\Ln \left \{ \frac{\left((B-1)^2-4 A C\right)}{A i} \right\} + \Ln (Ai )\right)+\frac{i 4q^2 A / \alpha  }{4 A C-(B-1)^2} 
	\right],
\label{MuNuIntegral}
\end{align}
where $\Ln(x)$ is the principal value of the logarithm with argument in the interval $(-\pi , \pi ]$ and the branch cut lies on the negative real axis.

Subsequently, the remaining integral over $\phi_i$ and $\lambda_i$ can be readily 
evaluated as follows: 
\begin{align}
\frac{1}{2\pi} \int_{0}^{\infty} d\lambda_k \int d \phi_k (1 + 2 i \phi_k /L)^{-L\alpha/2} 
\exp \left [
	-\frac{ a \phi_k^2 + 4 b \tau_k \phi_k - 4 c \tau_k}{2(1 + 2 i \phi_k /L)}  + i \phi_k \lambda_k
\right ] 
 =  T(a, b i, c,\tau_k) + \frac{1}{L} U(a,b i,c,\tau_k) + O(1/L^2),
\end{align}
where
\begin{align}
	T(a,b,c,\tau)=&\frac{1}{2} e^{-2 c \tau} \left(\text{erf}\left(\frac{\alpha +2 b\tau}{\sqrt{2a} }\right)+1\right), \nonumber \\
	U(a,b,c,\tau)=&-\frac{4 a c \tau+a+2 b \tau (\alpha +2 b\tau)}{\sqrt{2 \pi } a^{3/2}} \exp \left [-\frac{(\alpha +2 b\tau)^2}{2 a}-2 c \tau\right ].
	\label{TandU}
\end{align}
After summing over the $\tau_k$'s, we arrive at the equation
\begin{align}
	\NumMax =&\int_0^\infty da \int \frac{dA db dB dc dC}{(2\pi / L)^3 } \exp \left [
		\frac{U(a,bi,c,1) + U(a,bi,c,0)}{T(a,bi,c,1) + T(a,bi,c,0)}
	\right ]  \exp\left( L \Sigma(a, bi, c, Ai, B, Ci)\right)
	\label{TandUandSigma}
\end{align}
where 
\begin{align}
	\Sigma(a, b, c, A, B, C) = a A+b B+c C-\frac{1}{2} \alpha  \ln \left(4 A C+(B-1)^2\right) -\frac{4 A q^2}{4 A C+(B-1)^2} + \ln(T(a,b,c,1) + T(a,b,c,0)).
\label{Action}
\end{align}

The remaining integrals are hard to evaluate analytically.
Instead, we resort to the saddle-point method to obtain an asymptotic expansion of the integral.
Since $\Sigma$ is the exponential growth factor of the number of local maxima which must be a real number,
	one expects that the saddle points of \eqref{Action} are formed for the real arguments of $\Sigma$. 
This suggests that we should make the changes of variables $b \to b/i$, $A \to A/i$ and $C \to C/i$. 
For large $L$, the integrals are then dominated by the saddle point $(a^*, b^*, c^*, A^*, B^*, C^*)$ of $\Sigma(a, b, c, A, B, C)$.
If there is more than one saddle point, the one giving the largest value of $\Sigma(a, b, c, A, B, C)$ has to be chosen.
Then, the leading behavior of the number of maxima can be expressed in
terms of the saddle point as
\begin{align}
	\NumMax = \frac{1}{\sqrt{\vert \det H(\Sigma) \vert}} \exp\left[
		\frac{U(a^*,b^*,c^*,1) + U(a^*,b^*,c^*,0)}{T(a^*,b^*,c^*,1) + T(a^*,b^*,c^*,0)}
	\right] 	
	\exp\left( L \Sigma(a^*, b^*, c^*, A^*, B^*, C^*)\right),
	\label{NOptimaFinal}
\end{align}
where $H(\Sigma)$ is the Hessian matrix around the saddle point.
The reader may have noticed that the two principal values of the logarithm defined in \eqref{MuNuIntegral} are replaced by a real-valued logarithm in \eqref{Action}, which can be dangerous in general. 
However, it can be shown that this substitution is indeed correct by verifying that $(B^*-1)^2 + 4 A^*C^*$ is always positive for all saddle points of \eqref{Action}, and thus the imaginary arguments always cancel each other out.

Now, let us evaluate the saddle-point conditions.
The derivatives of $\Sigma$ with respect to $A$, $B$, $C$ are
\begin{align}
\frac{\partial \Sigma}{\partial A} &= a - 2 \frac{ \alpha C[ 4 A C + (B-1)^2] + 2 q^2 (B-1)^2}{[4AC + (B-1)^2]^2} , \qquad
\frac{\partial \Sigma}{\partial B} = b -  \frac{(B-1)[ \alpha (B-1)^2 + A ( 4 C \alpha - 8 q^2 )]}{[4AC + (B-1)^2]^2},\nonumber \\
\frac{\partial \Sigma}{\partial C} &= c -  \frac{2A[ \alpha (B-1)^2 + A ( 4 C \alpha - 8 q^2 )]}{[4AC + (B-1)^2]^2} .
\label{Eq:ABC}
\end{align}
By requiring that the above three equations are zero at the saddle point, 
we get
\begin{align}
	A=\frac{\alpha  c}{2 \left(a c+b^2\right)},\quad B-1 = \frac{\alpha  b}{a c+b^2},\quad C = 
	\frac{1}{4} \left(\frac{2 a \alpha }{a c+b^2}+\frac  {-\alpha \pm \sqrt{\alpha ^2-16 c q^2}}{c}\right).
	\label{ABCSolution}
\end{align}
The two solutions of $C$ force us to perform a two-fold analysis for the remaining integrals since we cannot \textit{a priori} determine which solution will yield the correct saddle point.
Instead, we introduce another real number $ g = \pm \sqrt{\alpha ^2-16 c q^2}$ which is allowed to take both signs. Then, by imposing the functional relation $c(g) = (\alpha^2 - g^2)/(16 q^2)$, both solutions are covered by a single analysis. 
In this way, the saddle point is obtained in terms of $g$ instead of $c$.
Finally, substituting this solution into \eqref{NOptimaFinal} gives \eqref{ReducedComplexity}.

\section{\label{appendixFB}Mean Phenotypic Distance \boldmath{$z^*$} in the Joint Limit}
In this appendix, we will associate
the saddle-point value $a^*$ of
the variable $a$ entering the complexity function \eqref{ReducedComplexity} 
with the mean phenotypic distance $z^\ast$. 
To this end, we first consider the probability density $P(\tau,a)$
that a genotype $\tau$ whose phenotypic vector is of squared magnitude $L^2a/4$ is a local maximum. Formally, we can write
\begin{align}
P(\tau,a) &= \int D\xi \prod_k \theta(\mathcal{E}_k)
\delta \left ( a - \frac{4}{L^2} \left (\vec{Q} + \sum_k \vec{\xi}_k \tau_k\right )^2
\right ) 
= \int \frac{L}{2\pi}dA \int D\xi \prod_k \theta(\mathcal{E}_k) 
\exp  \left [
	i L a A - i \frac{4A}{L} \left (\vec{Q} + \sum_k \vec{\xi}_k \tau_k \right )^2
\right ]\nonumber\\
&= \frac{L}{2\pi} \int d A \int D\xi \prod_k \theta(\mathcal{E}_k) \int \mathcal{D} \psi \exp  \left[
	i L a A + i \frac{L}{16A}\vec{\psi}^2 +  i \vec{\psi } \cdot \left ( \vec{Q} + \sum_k \vec{\xi}_k \tau_k \right )
\right ],
\end{align}
where we have used the identity $\int_{-\infty}^\infty dx \exp(i  p x^2 + i q x)
= \sqrt{\pi/p} e^{-i\pi/4} \exp[-i q^2/(4p)]$ for $p>0$, $ \mathcal{D}
\psi = \prod_s (\sqrt{L} e^{i\pi/4} d\psi_s)/(4 \sqrt{\pi  A}) $, and
the notation is the same as in \ref{appendixF:JoinLimit}. 
Following the same procedure in the previous appendix, we get
\begin{align}
P(\tau,a)=& \int \Measure{\xi} \Measure{\lambda} \Measure{\phi} \Measure{\mu} \Measure{\nu} e^{i \phi \cdot \lambda + i L \vec{\mu}\cdot\vec{\nu}}\exp \left[
\sum_k \frac{i}{L} \phi_k \left(2 \vec{\xi}_k \cdot \vec{Q}  - |\vec{\xi}_k|^2\right)
+ i \sum_{k} \vec{\xi}_k \cdot (- \phi_k \vec{\mu} + 2\tau_k \vec{\nu})
\right]  \nonumber \\
&\times \frac{L}{2\pi} \int d A  \int \mathcal{D} \psi \exp
\left [	i L a A + i \frac{L}{16A}\vec{\psi}^2 +  i \vec{\psi } \cdot ( \vec{Q} + \sum_i \vec{\xi}_i \tau_i)
\right ].
\end{align}
By shifting $\vec{\nu} \to \vec{\nu} - \vec{\psi}/2$ and integrating over $\vec{\psi}$, we have
\begin{align}
\label{Ptaua}
P(\tau,a) =& \int \Measure{\xi} \Measure{\lambda} \Measure{\phi} \Measure{\mu} \Measure{\nu} e^{i \phi \cdot \lambda + i L \vec{\mu}\cdot\vec{\nu}}\exp \left[
\sum_k \frac{i}{L} \phi_k \left(2 \vec{\xi}_k \cdot \vec{Q}  - |\vec{\xi}_k|^2\right)
+ i \sum_{k} \vec{\xi}_k \cdot (- \phi_k \vec{\mu} + 2\tau_k \vec{\nu})
\right]  \nonumber \\
&\times \frac{L}{2\pi} \int d A  \int \mathcal{D} \psi \exp
\left [
i L a A + i \frac{L}{16A}\vec{\psi}^2 +  i \vec{\psi } \cdot ( \vec{Q} - L \vec{\mu}/2)
\right ]\nonumber \\
=&\int \Measure{\xi} \Measure{\lambda} \Measure{\phi} \Measure{\mu} \Measure{\nu} e^{i \phi \cdot \lambda + i L \vec{\mu}\cdot\vec{\nu}}\exp \left[
\sum_k \frac{i}{L} \phi_k \left(2 \vec{\xi}_k \cdot \vec{Q}  - |\vec{\xi}_k|^2\right)
+ i \sum_{k} \vec{\xi}_k \cdot (- \phi_k \vec{\mu} + 2\tau_k \vec{\nu})
\right] \nonumber\\
&\times \int \frac{d A }{2\pi/L} \exp  \left\{ i L A
	\left [ a  - \left (\frac{2\vec{Q}}{L} - \vec{\mu}
\right )^2 \right ]
\right\}.
\end{align}
Since we have set $Q_s = Q/\sqrt{n}$ for $s=1,\ldots,n$, 
the last integral becomes 
\begin{align}
\int \frac{d A}{2\pi/L} \exp  \left\{ i L A
	\left [ a  - \left (\frac{2\vec{Q}}{L} - \vec{\mu}
\right )^2 \right ] \right \}
	=\int \frac{d A }{2\pi/L} \exp  
\left\{ i L A \left [ a -  \sum_s  (\mu_s - \rr)^2 \right ]
	\right\}.
\end{align}
Since $\NumMax = \Tr \int_0^\infty da P(\tau,a)$, by applying the
manipulations of \ref{appendixF:JoinLimit} to \eqref{Ptaua}
we arrive at the same integral form as in \eqref{LagrangeMultipliers}. 
Since $\sum_\tau P(\tau,a)$ is the mean number of local maxima 
whose phenotypic vectors have squared magnitude $a$, 
we see that the saddle point $a^*$ of \eqref{Action} 
determines the mean phenotypic distance $z^\ast$ 
through 
\begin{align}
	z^* = L \frac{\sqrt{a^*}}{2}.
	\label{RelationAandZ}
\end{align}
This shows in particular that $z^*$ is linear in $L$. 

\section{\label{appendix:AFB}Mean Genotypic Distance \boldmath{$\rho^*$} in the Joint Limit}

To have access to the information about the typical value of the 
genotypic (Hamming) distance of a local fitness maximum from the wild type,
we rewrite \eqref{TandUandSigma} as
\begin{align}
	\NumMax =& \int_0^\infty da \int \frac{dA db dB dc dC}{(2\pi / L)^3 } \sum_{s=0}^{L} \binom{L}{s}  \left[
		T(a,b,c,1) + \frac{U(a,b,c,1)}{L}
	\right]^s
	\left[
		T(a,b,c,0) + \frac{U(a,b,c,0)}{L}
	\right]^{L-s} \nonumber \\
	\approx &	\int_0^\infty da \int \frac{dA db dB dc dC}{(2\pi / L)^3 } \int_{0}^{1} d\rho  
		\frac{ e^{L \Sigma(a,b,c,A,B,C,\rho) }} {\sqrt{2 \pi  L \rho(1-\rho)  }}
		\exp\left\{
			\rho \left[1 + \frac{U(a,b,c,1)}{T(a,b,c,1)}\right] + (1-\rho) \left[1 + \frac{U(a,b,c,0)}{T(a,b,c,0)}\right]
		\right\}, 
\end{align}
where we have rearranged the summation $\sum_\tau$ as
$\sum_{s=0}^L \binom{L}{s}$ taking advantage of the inherent
permutation symmetry, Stirling's formula has been used to evaluate the
binomial coefficients, $\sum_s$ is approximated as $ L \int_0^1 d\rho$ with $s = L\rho$, and
\begin{align}
	\Sigma(a, b, c, A, B, C, \rho) \equiv & a A+b B+c C-\frac{1}{2} \alpha  \ln \left[4 A C+(B-1)^2\right] -\frac{4 A q^2}{4 A C+(B-1)^2} \nonumber\\
	&+  \rho \ln T(a,b,c,1) + (1-\rho) \ln T(a,b,c,0) - \rho \ln \rho - (1-\rho)\ln(1-\rho).
\label{Action2}
\end{align}
The saddle-point equations for this expression involve seven variables including $\rho$.
Since the saddle-point equations for $A$, $B$, $C$ are the same
as \eqref{Eq:ABC}, we may again
insert \eqref{ABCSolution} into \eqref{Action2}, which yields 
\begin{align}
	\Sigma^{\mathrm{red}}(a,b,g,\rho) =&-\ln 2 + \frac{\alpha}{2} 
\left ( 1 - \ln \frac{\alpha}{2} \right ) -\frac{\alpha}{2}\ln \left[ \frac{\alpha +g}{a c(g)+b^2}\right]+b+\frac{g}{2}
	-\rho  \ln \rho  -(1-\rho)\ln (1-\rho) \nonumber \\
	 &
	+(1-\rho) \ln \left(\text{erf}\left(\frac{\alpha }{\sqrt{2a}}\right)+1\right) 
	+\rho  \left\{\ln \left[\text{erf}\left(\frac{\alpha +2 b}{\sqrt{2a} }\right)+1\right]-2 c(g)\right\}.
	\label{ReducedComplexityRho}
\end{align}
Since 
\begin{align}
\frac{\partial \Sigma}{\partial \rho} = \ln \frac{T(a,b,c,1)}{T(a,b,c,0)}
- \ln \rho + \ln (1-\rho),
\end{align}
the saddle-point value of $\rho^*$ is
\begin{align}
\rho^* = \frac{T(a^*,b^*,c^*,1)}{T(a^*,b^*,c^*,1)+T(a^*,b^*,c^*,0)}=\left \{
1 + e^{2c^*} \left [ \frac{\text{erf}\left ( \alpha/\sqrt{2 a^*} \right ) + 1}{\text{erf}\left ( (\alpha + 2 b^*)/\sqrt{2 a^*} \right ) + 1}
\right ] \right \}^{-1}.
\label{RhoSolution}
\end{align}
By inserting $\rho^*$ into the saddle-point equations for $a, b, c$,
one can easily see that the final equations are the same as those
derived from \eqref{Action}.

\section{Derivation of \eqref{SaddlePointThird}}
\label{appendix:ThirdSaddlePoint}
The determination of the solution describing regime III relies on the
intuition that as $\alpha$ becomes large, the fitness
landscape is asymptotically linear with the wild type being 
the global fitness maximum, as demonstrated in \textit{Sign epistasis}
for $L=2$.
This suggests an ansatz where $a^*$ is close to $4q^2$, which corresponds to the wild-type phenotypic distance as shown in \eqref{RelationAandZ}. 
Given this clue, one can additionally find that
\begin{align}
\frac{\partial}{\partial  a} \Sigma^{\mathrm{red}}(a, b, g) = 0
\end{align}
is solved by $a = 4q^2$, $b = -\alpha$ and $g = \alpha$.
Furthermore, if we evaluate the remaining saddle-point conditions around this point, we find that this solution fails to solve them by a slight margin,
\begin{align}
\left. \frac{\partial}{\partial  b} \Sigma^{\mathrm{red}}(a, b, g) \right|_{a = 4q^2,\,  b = -\alpha, \, g = \alpha} = \frac{e^{-\frac{\alpha ^2}{8 q^2}}}{\sqrt{2 \pi } q}
\end{align}
and 
\begin{align}
\left. \frac{\partial}{\partial  c} \Sigma^{\mathrm{red}}(a, b, g) \right|_{a = 4q^2,\,  b = -\alpha, \, g = \alpha} = \frac{\alpha }{8 q^2}  \text{erfc}\left(\frac{\alpha }{2 \sqrt{2} q}\right).
\end{align}
Given the fact that $  \text{erfc}(x) = e^{-x^2} \left[1/(\sqrt{\pi }
	x)+O\left(x^{-3}\right)\right] $, these
    non-vanishing terms are seen to be of the order of $\epsilon = e^{-\alpha ^2/(8 q^2)}$.
Hence, it is sufficient to consider an expansion around the zeroth-order solution of the form $ \Sigma^{\mathrm{red}}(4q^2 + A_1 \epsilon ,-\alpha + A_2 \epsilon,\alpha + A_3 \epsilon)  $ 
to show that \eqref{SaddlePointThird} satisfies the saddle-point conditions \eqref{ReducedComplexityExtremumCondition}. 
To this end, we first focus on the derivatives with respect to $A_1$ and $A_2$,
\begin{align}
	\frac{1}{\epsilon} \frac{\partial}{\partial  A_1} \Sigma^{\mathrm{red}}(4q^2 + A_1 \epsilon ,-\alpha + A_2 \epsilon,\alpha + A_3 \epsilon) &= -\frac{A_3 \epsilon }{16 q^2} + O(\epsilon^2), \nonumber \\ 
	\frac{1}{\epsilon} \frac{\partial}{\partial  A_2} \Sigma^{\mathrm{red}}(4q^2 + A_1 \epsilon ,-\alpha + A_2 \epsilon,\alpha + A_3 \epsilon) &=   \left(\frac{1}{\sqrt{2 \pi } q}-\frac{2A_2 + A_3}{2\alpha }\right) \epsilon + O(\epsilon^2).
\end{align}
The vanishing contributions in $\epsilon$ imply that the zeroth-order solution $ (4q^2, -\alpha, \alpha) $ satisfies the first two saddle point conditions.
Additionally, we find that the corrections of the order $O(\epsilon)$ are $A_3 =0$ and $A_2 = \alpha/(\sqrt{2\pi}q)$.
Since $A_3=0$ to leading order, the saddle point equation with respect
to $g$ should be evaluated to order $O(\epsilon^2)$. This yields
\begin{align}
	\frac{1}{\epsilon^2} \frac{\partial}{\partial  B_3} \Sigma^{\mathrm{red}}(4q^2 + A_1 \epsilon ,-\alpha + A_2 \epsilon,\alpha + B_3 \epsilon^2) &=  \left[
		 -\frac{A_1  }{16 q^2}-\frac{\sqrt{\frac{2}{\pi }} q  }{\alpha ^2} + O\left(\frac{q^3}{\alpha^4}\right)
	\right] \epsilon  + O(\epsilon^2),
\end{align}
and subsequently, $A_1$ is solved to be $A_1 = \left[
(16 \sqrt{2/\pi} q^3/\alpha ^2) + O(q^4/\alpha^3)
\right] \epsilon + O(\epsilon^2)$.
Finally, by inserting the solutions $A_1$, $A_2$ and $A_3$ as well as the zeroth-order solutions into \eqref{RhoSolution}, the solution for $\rho^*$ is found to be
\begin{align}
	\rho^* = \left[
		\frac{\sqrt{\frac{2}{\pi }} q \epsilon }{\alpha }  + O\left(\frac{q^3}{\alpha^3}\right)
	\right]+  O(\epsilon^2).
\end{align}

\end{document}